\documentclass[fleqn,12pt,twoside]{article}
\usepackage{espcrc1}
\usepackage{graphicx}
\usepackage[figuresright]{rotating}

\newcommand{\AmS}{{\protect\the\textfont2
  A\kern-.1667em\lower.5ex\hbox{M}\kern-.125emS}}
\newcommand{\eq}{{\,=\,}}

\title{Early thermalization at RHIC\thanks{This work was supported by 
               the U.S. Department of Energy
               under contract DE-FG02-01ER41190.}}

\author{Ulrich Heinz\thanks{Invited speaker} and Peter Kolb\\[1ex]
        Department of Physics, The Ohio State University, Columbus, 
        OH 43210, USA}

\begin{document}

% typeset front matter
\maketitle

\begin{abstract}
{
%\small 
It is shown that recent RHIC data on hadron spectra and elliptic 
flow can be excellently reproduced within a hydrodynamic description of
the collision dynamics, and that this provides strong evidence for rapid 
thermalization while the system is still in the quark-gluon plasma phase. 
But even though the hydrodynamic approach provides an impressive 
description of the single-particle momentum distributions, it fails to 
describe the two-particle momentum correlation (HBT) data for central Au+Au
collisions at RHIC. We suggest that this is not likely to be repaired by 
further improvements in our understanding of the early collision stages, 
but probably requires a better modelling of the freeze-out process. We 
close with a prediction of the phases of the azimuthal oscillations of 
the HBT radii in noncentral collisions at RHIC.
}
\end{abstract}

%%%%%%%%%%%%%%%%%%%%%%%%%%%%%%%%%%%%%%%%%%%%%%%%%%%%%%%%%%%%%%%%%%%%%%%%
\section{ELLIPTIC FLOW AS AN EARLY QGP SIGNATURE}
\label{sec1}
%%%%%%%%%%%%%%%%%%%%%%%%%%%%%%%%%%%%%%%%%%%%%%%%%%%%%%%%%%%%%%%%%%%%%%%%

The quark-gluon plasma (QGP) is a thermalized system and, as such, has 
thermal pressure. If the QGP is created in a heavy-ion collision, this 
pressure acts against the surrounding vacuum and
causes a rapid collective expansion (``flow'') of the reaction zone,
the ``Little Bang''. Collective flow is an unavoidable consequence of
QGP formation in heavy-ion collisions, and its absence could be taken as
proof that no such plasma was ever formed. Its presence, on the other hand,
does not automatically signal QGP formation. Detailed studies of the 
observed final state flow pattern are necessary to convince oneself that 
the reflected time-integrated pressure history of the collision region 
indeed requires a thermalized state in the early collision stage whose 
pressure and energy density are so high that it can no longer be 
mistaken as consisting of conventional hadronic matter. 

Much progress in this direction was recently achieved by studying {\em 
elliptic flow} \cite{O92} in non-central (non-zero impact parameter) 
heavy-ion collisions. It characterizes the azimuthal anisotropy in the 
transverse plane of the final momentum distribution near midrapidity 
($y\eq0$) and is quantified by the second harmonic coefficient 
$v_2(y,p_\perp;b)$ of a Fourier expansion in $\phi_p$ of the measured 
hadron spectrum $dN/(dy\,p_\perp dp_\perp\,d\phi_p)$ \cite{VZ96}. (The
first harmonic coefficient $v_1$ measures the {\em directed flow} or 
``bounce-off'' of the colliding nuclei at forward and backward rapidities
\cite{Stocker:1986ci}; it vanishes at midrapidity by symmetry.)

Since individual nucleon-nucleon collisions produce azimuthally symmetric
spectra, such final state momentum anisotropies must be generated dynamically 
during the nuclear reaction. They require the existence of an initial spatial 
anisotropy of the reaction zone, either by colliding deformed nuclei 
\cite{Sh00,Li00,Kolb:2000sd} or by colliding spherical nuclei at non-zero 
impact parameter $b{\,\ne\,}0$ (the practical method of choice so far). 
The transfer of the initial spatial anisotropy onto a final momentum 
anisotropy requires final state interactions (rescattering) within
the produced matter; without them the primordial azimuthally symmetric 
momentum distribution survives. Microscopic transport calculations 
\cite{Zhang:1999rs,Molnar:2001ux} show a monotonic dependence of $v_2$ 
on the opacity (density times scattering cross section) of the produced 
matter which is inversely related to its thermalization time. These studies
strongly suggest that, for a given initial spatial anisotropy $\epsilon_x$, 
the maximum momentum-space response $v_2$ is obtained in the {\em hydrodynamic 
limit} which assumes perfect local thermal equilibrium at every space-time 
point (i.e. a thermalization time which is much shorter than any macroscopic 
time scale in the system). Any significant delay of thermalization (modelled,
for example, as an initial free-streaming stage) causes a decrease of the 
initial spatial anisotropy without concurrent build-up of momentum 
anisotropies, thereby reducing the finally observed elliptic flow signal
\cite{Kolb:2000sd}.

This specific sensitivity of the elliptic flow to rescattering and 
pressure build-up in the early collision stages 
\cite{Sorge:1997pc,Voloshin:2000gs} (before the spatial deformation and 
the resulting anisotropies of the pressure gradients have disappeared 
\cite{Kolb:2000sd}) puts $v_2$ on the list of ``early signatures'' of 
the collision dynamics. In contrast to other early probes (which use 
rare signals such as hard photons and dileptons,
heavy quarkonia and jets), $v_2$ can be extracted from the bulk of
the measured hadrons which are very abundant and thus easily accessible.
In fact, the elliptic flow measurement in Au+Au collisions at 
$\sqrt{s}\eq130\,A$\,GeV \cite{Ackermann:2001tr} became the {\em second} 
publication of RHIC data and appeared within days of the end of the first 
RHIC run.

In this talk I will present results from hydrodynamic simulations of
hadronic spectra and elliptic flow at RHIC energies. We will see that
the hydrodynamic approach provides an excellent quantitative description 
of the bulk of the data and fails only for very peripheral Au+Au collisions
and/or at high $p_\perp{>}1.5{-}2$\,GeV/$c$. That the hydrodynamic approach
fails if the initial nuclear overlap region becomes too small or the 
transverse momentum of the measured hadrons becomes too large is not 
unexpected. However, where exactly hydrodynamics begins to break
down gives important information about the microscopic rescattering
dynamics. What is really surprising is that the hydrodynamic approach
works so well in semi-central collisions where it quantitatively reproduces
the momenta of more than 99\% of the particles: below $p_\perp\eq1.5$\,GeV/$c$
the elliptic flow data \cite{Ackermann:2001tr,Lacey:2001va,Poskanzer:2001cx} 
actually exhaust the hydrodynamically predicted 
\cite{Kolb:2000sd,Kolb:1999it,Teaney:2001cw,Kolb:2001fh,Huovinen:2001cy} 
upper limit. The significance of this agreement can hardly be overstressed 
(see also E. Shuryak's concluding talk), and it poses significant challenges
for microscopic descriptions of the early collision dynamics. 

However, not all is well with hydrodynamics: in its present form,
it fails to reproduce the HBT measurements \cite{Adler:2001zd,Johnson:2001zi}
which constrain the freeze-out distribution in space-time. This problem, 
which affects more our understanding of the late freeze-out stage than 
of the early thermalization stage, will be discussed in Sec.~\ref{sec4}.

%%%%%%%%%%%%%%%%%%%%%%%%%%%%%%%%%%%%%%%%%%%%%%%%%%%%%%%%%%%%%%%%%%%%%%%%%%%%%
\section{HYDRODYNAMIC FIREBALL EXPANSION}
\label{sec2}
%%%%%%%%%%%%%%%%%%%%%%%%%%%%%%%%%%%%%%%%%%%%%%%%%%%%%%%%%%%%%%%%%%%%%%%%%%%%%

The natural language for describing collective flow phenomena is 
hydrodynamics. Its equations control the space-time evolution of
the pressure, energy and particle densities and of the local fluid 
velocity. The system of hydrodynamic equations is closed by specifying
an {\em equation of state} which gives the pressure as a function of 
the energy and particle densities. In the ideal fluid (non-viscous) 
limit, the approach assumes that the microscopic momentum distribution 
is thermal at every point in space and time (note that this does not 
require chemical equilibrium -- chemically non-equilibrated situations 
can be treated by solving separate and coupled conservation equations 
for the particle currents of individual particle species). Small 
deviations from local thermal equilibrium can in principle be dealt 
with by including viscosity, heat conduction and diffusion effects, 
but such a program is made difficult in practice by a number of 
technical and conceptual questions \cite{Rischke:1998fq} and has so 
far not been very successful for relativistic fluids. Stronger 
deviations from local thermal equilibrium require a microscopic 
phase-space approach (kinetic transport theory), but in this case the 
concepts of an equation of state and of a local fluid velocity field 
themselves become ambiguous, and the direct connection between flow 
observables and the equation of state of the expanding matter is lost. 

The assumption of local thermal equilibrium in hydrodynamics is an 
external input, and hydrodynamics offers no insights about the 
equilibration mechanisms. It is clearly invalid during the initial 
particle production and early recattering stage, and it again breaks 
down towards the end when the matter has become so dilute that 
rescattering ceases and the hadrons ``freeze out''. The hydrodynamic 
approach thus requires a set of {\em initial conditions} for the 
hydrodynamic variables at the earliest time at which the assumption of
local thermal equilibrium is applicable, and a {\em ``freeze-out 
prescription''} at the end. For the latter we use the Cooper-Frye algorithm 
\cite{Cooper:1974mv} which implements an idealized sudden transition 
from perfect local thermal equilibrium to free-streaming. This is not 
unreasonable because freeze-out (of particle species $i$) is controlled 
by a competition between the local expansion rate $\partial\cdot u(x)$ 
(where $u^\mu(x)$ is the fluid velocity field) and the local scattering 
rate $\sum_j \langle \sigma_{ij} v_{ij}\rangle \rho_j(x)$ (where the sum
goes over all particle species with densities $\rho_j(x)$ and 
$\langle \sigma_{ij} v_{ij}\rangle$ is the momentum-averaged transport 
cross section for scattering between particle species $i$ and $j$, 
weighted with their relative velocity); while the local expansion rate 
turns out to have a rather weak time-dependence, the scattering
rate drops very steeply as a function of time, due to the rapid
dilution of the particle densities $\rho_j$ \cite{Schnedermann:1994gc},
causing a rapid transition to free-streaming. -- A better algorithm 
\cite{Bass:2000ib,Teaney:2001cw} switches from a hydrodynamic description 
to a microscopic hadron cascade at or shortly after the quark-hadron 
transition, before the matter becomes too dilute, and lets the cascade 
handle the freeze-out kinetics. The resulting flow
patterns \cite{Teaney:2001cw} from such an improved freeze-out algorithm 
don't differ much from our simpler Cooper-Frye based approach. 

The main advantage of the microscopic freeze-out algorithm 
\cite{Bass:2000ib,Teaney:2001cw} is that it also correctly reproduces 
the final chemical composition of the fireball, since the particle 
abundances already freeze out at hadronization, due to a lack of 
particle-number changing inelastic rescattering processes in the 
hadronic phase \cite{Heinz:1999kb}. Our version of the hydrodynamic 
approach uses an equation of state which assumes local chemical 
equilibrium all the way down to kinetic freeze-out at 
$T_{\rm f}{\,\approx\,}125$\,MeV and thus is unable to reproduce the 
correct hadron yield ratios. We therefore adjust the normalization of 
the momentum spectra for the rarer particle species (kaons, protons, 
antiprotons) by hand to reproduce the chemical equilibrium ratios at a 
chemical freeze-out temperature $T_{\rm chem}\eq165$\,MeV. The absolute 
normalization of the 
pion spectra is adjusted through the initial energy density in central 
collisions; for non-central collisions no new parameters enter since 
the centrality dependence of the initial conditions is completely 
controlled by the collision geometry (see below).

%%%%%%%%%%%%%%%%%%%%%%%%%%%%%%%%%%%%%%%%%%%%%%%%%%%%%%%%%%%%%%%%%%%%%%%%%%%%%
\section{RADIAL AND ELLIPTIC FLOW FROM HYDRODYNAMICS}
\label{sec3}
%%%%%%%%%%%%%%%%%%%%%%%%%%%%%%%%%%%%%%%%%%%%%%%%%%%%%%%%%%%%%%%%%%%%%%%%%%%%%

We solve the relativistic equations for ideal hydrodynamics (for 
details see \cite{Kolb:1999it,Kolb:2000sd}). To simplify the numerical 
task, we analytically impose boost invariant longitudinal expansion 
\cite{Bjorken:1983qr,O92}, without giving up any essential physics as 
long as we focus on the transverse ex\-pan\-sion dynamics near midrapidity 
(the region which most RHIC experiments cover best). 

%%%%%%%%%%%%%%%%%%%%%%%%% Fig. 1 %%%%%%%%%%%%%%%%%%%%%%%%%%%%%%%%%%%%%%%%%%%%
\vspace*{-3mm}
\begin{figure}[htb]
\begin{minipage}[t]{52mm}
\includegraphics[height=50mm,width=53mm]{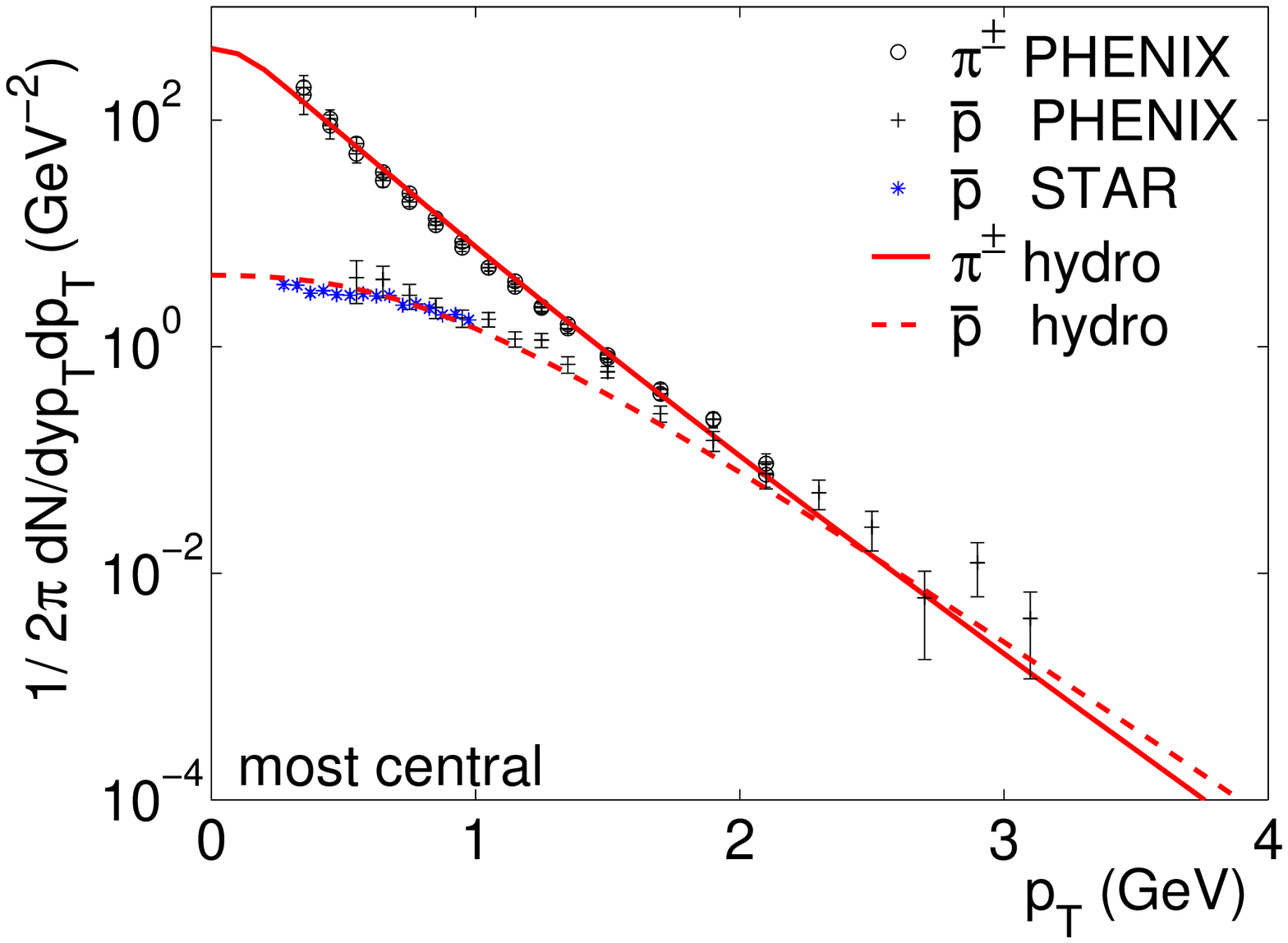}
%\caption{Good sharp prints should be used and not (distorted) photocopies.}
\end{minipage}
\begin{minipage}[t]{52mm}
\vspace*{-49.9mm}
\includegraphics[height=50mm,width=53mm]{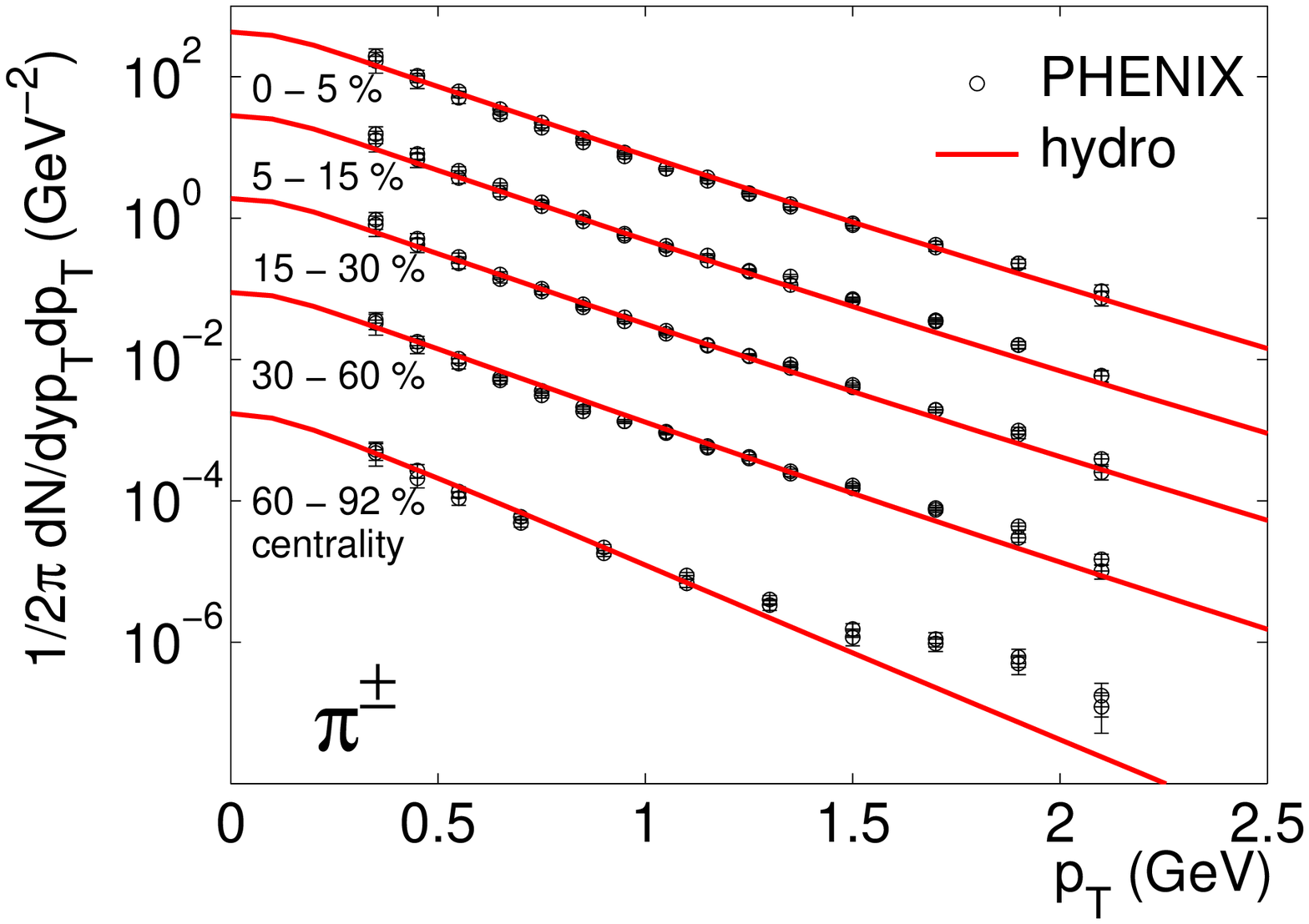}
\end{minipage}
%\hspace{\fill}
%
\begin{minipage}[t]{52mm}
\vspace*{-50mm}\hspace*{-1mm}
\includegraphics[height=50.5mm,width=53mm]{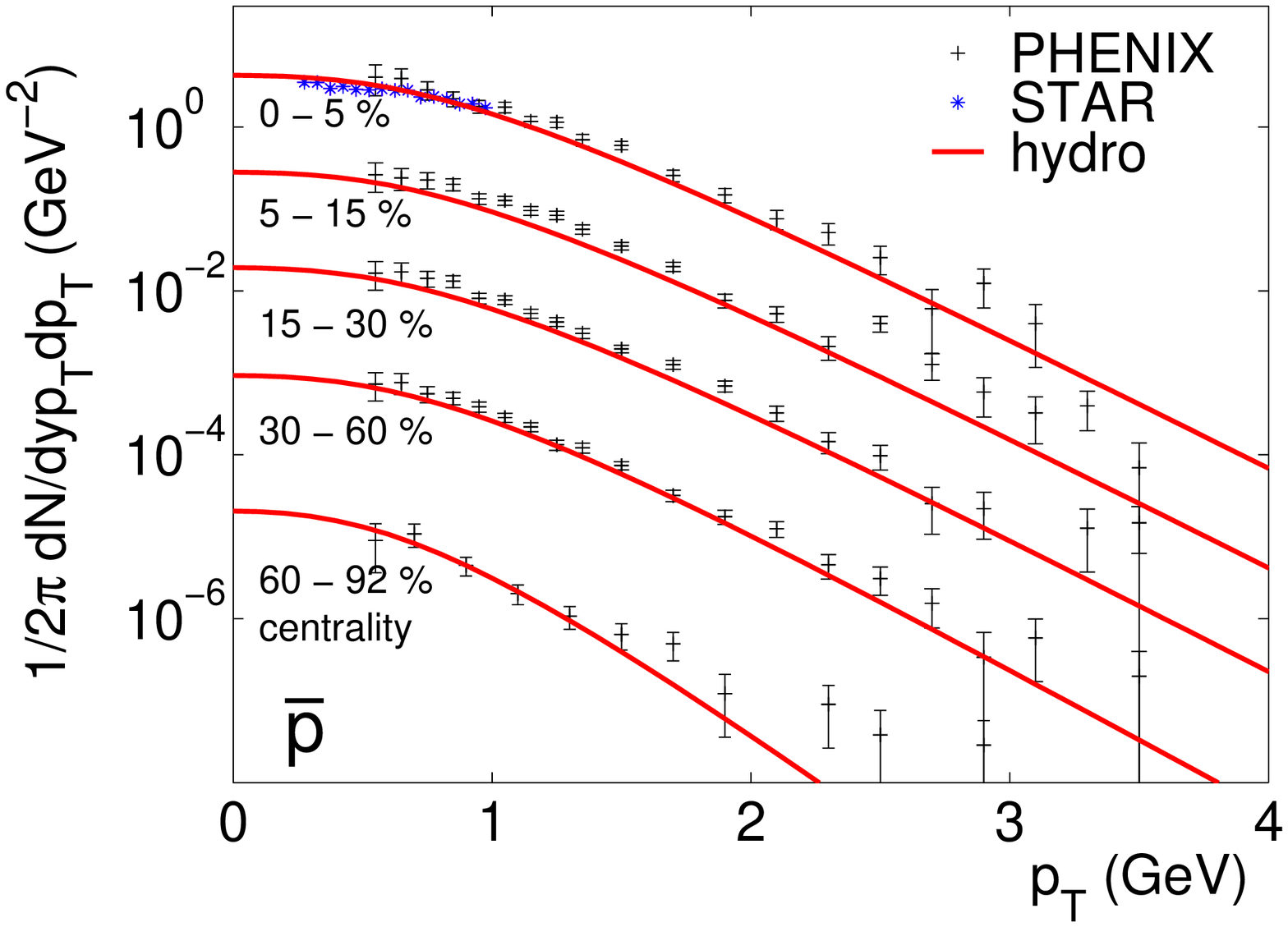}
\end{minipage}
\vspace*{-15mm}
\begin{center}
\begin{minipage}[t]{150mm}
\caption{\label{F1} 
\small
Charged pion and antiproton spectra from central (left
panel) and semi-central to peripheral (middle and right panel) Au+Au 
collisions at $\sqrt{s}\eq130\,A$\,GeV. The data were taken by the PHENIX 
\cite{PHENIX_spec} and STAR \cite{STAR_spec} collaborations, the curves
show hydrodynamical calculations (for details see text).}
\end{minipage}
\end{center}
\end{figure}
%%%%%%%%%%%%%%%%%%%%%%%%%%%%%%%%%%%%%%%%%%%%%%%%%%%%%%%%%%%%%%%%%%%%%%%%%%%%%%%

\vspace*{-11mm}

The hydrodynamic expansion starts at time $\tau_{\rm eq}$ which we 
fixed by a fit to hadron spectra from central Pb+Pb collisions at the 
SPS and then extrapolated to RHIC initial conditions, taking into account 
the higher initial parton density \cite{Kolb:2000sd}. For each impact
parameter the initial energy density profile in the transverse plane
is calculated from a Glauber parametrization using a realistic nuclear 
thickness function \cite{Kolb:2000sd,KHHET}. We have explored a number 
of different options involving various combinations of ``hard'' and 
``soft'' production mechanisms for the initial particle production 
\cite{KHHET}. While all were found to give almost indistinguishable 
results for the charged particle elliptic flow, the radial flow (i.e. 
the slopes of the hadron transverse mass spectra) and the centrality 
dependence of the charged particle rapidity density at midrapidity, 
$(dN_{\rm ch}/dy)(y\eq0)$, exhibit sensitivity to the initial transverse 
density profile \cite{KHHET}. We here present results for initial 
conditions at $\tau_{\rm eq}\eq0.6$\,fm/$c$ calculated from a mixture of 
25\% ``hard'' (binary collision) and 75\% ``soft'' (wounded nucleon) 
contributions \cite{KHHET} to the initial {\em entropy density} (or 
parton density), with a maximal entropy density $s_{\rm max}\eq85/$fm$^3$
at the fireball center in central collisions (corresponding to a maximal 
energy density $e_{\rm max}\eq21.4$\,GeV/fm$^3$ and a maximal temperature 
$T_{\rm max}\eq328$\,MeV). At the standard time $\tau\eq1$\,fm/$c$ for energy 
density estimates from the measured multiplicity density using Bjorken's
formula \cite{Bjorken:1983qr}, this corresponds to an average energy
density $\langle e\rangle(1\,{\rm fm}/c)\eq5.4$\,GeV/fm$^3$ which is about
70\% higher than the value reported from 158\,$A$\,GeV Pb+Pb collisions
at the SPS. (Note that $\langle e\rangle$ at $\tau_{\rm eq}\eq0.6$\,fm/$c$ 
is nearly twice as large!) The corresponding central values and profiles 
for peripheral collisions are then given by the Glauber model 
\cite{Kolb:2000sd,KHHET}. For the kinetic freeze-out temperature we
took $T_{\rm f}\eq128$\,MeV, independent of centrality.

In Fig.~\ref{F1} we show the (absolutely normalized) single particle 
$p_\perp$-spectra for charged pions and antiprotons measured in Au+Au 
collisions at RHIC together with the hydrodynamical results. The latter
were normalized in central collisions as described at the end of 
Sec.~\ref{sec2}, but their centrality dependence and shapes are then 
completely fixed by the model. The agreement with the data is impressive;
for antiprotons the data go out to $p_\perp{\,\leq\,}3$\,GeV/$c$, and the 
hydrodynamic model still works within error bars! Only for very peripheral 
collisions (impact parameter $b{\,>\,}10$\,fm) the data show a significant 
excess of high-$p_\perp$ particles at $p_\perp{\,>\,}1.5$\,GeV/$c$. Teaney 
{\em et al.} \cite{Teaney:2001cw} showed that this excellent agreement 
requires a phase transition (soft region) in the equation of state;
without the transition, the agreement is lost, especially when the 
constraints from SPS data and from the elliptic flow measurements below
are taken into account.

It should be stressed that in the hydrodynamic picture the fact that
for $p_\perp{\,>\,}2$\,GeV/$c$ antiprotons become more abundant than
pions (left panel of Fig.~\ref{F1}) is not surprising, but a simple 
consequence of the strong radial flow at RHIC. For a hydrodynamically
expanding thermalized fireball, at relativistic transverse momenta 
$p_\perp{\,\gg\,}m_0$ all hadron spectra have the same slope 
\cite{Lee:1990sk}, and at fixed $m_\perp{\,\gg\,}m_0$ their relative
normalization is given by $(g_i\lambda_i)/(g_j\lambda_j)$ (where $g_{i,j}$
is the spin-isospin degeneracy factor and $\lambda_{i,j}=e^{\mu_{i,j}/T}$ 
is the fugacity of hadron species $i,j$). At RHIC the baryon chemical 
potential at chemical freeze-out is small, $\mu_B/T_{\rm
 chem}{\,\approx\,}0.26$ \cite{Braun-Munzinger:2001ip}, and $\mu_\pi\eq0$; 
the $\bar p/\pi^-$ ratio at fixed and sufficiently large $m_\perp$ is thus 
{\em predicted} to be larger than 1: $(\bar p/\pi^-)_{m_\perp}\eq2\exp[
 -(\mu_B{+}\mu_\pi)/T_{\rm chem}]{\,\approx\,}1.5$ (where the factor 2 
arises from the spin degeneracy of the $\bar p$).

%%%%%%%%%%%%%%%%%%%%%%%%% Fig. 2 %%%%%%%%%%%%%%%%%%%%%%%%%%%%%%%%%%%%%%%%%%%%
\vspace*{-5mm}
\begin{figure}[htb]
\begin{minipage}[t]{52mm}
\includegraphics*[height=50mm,width=51mm]{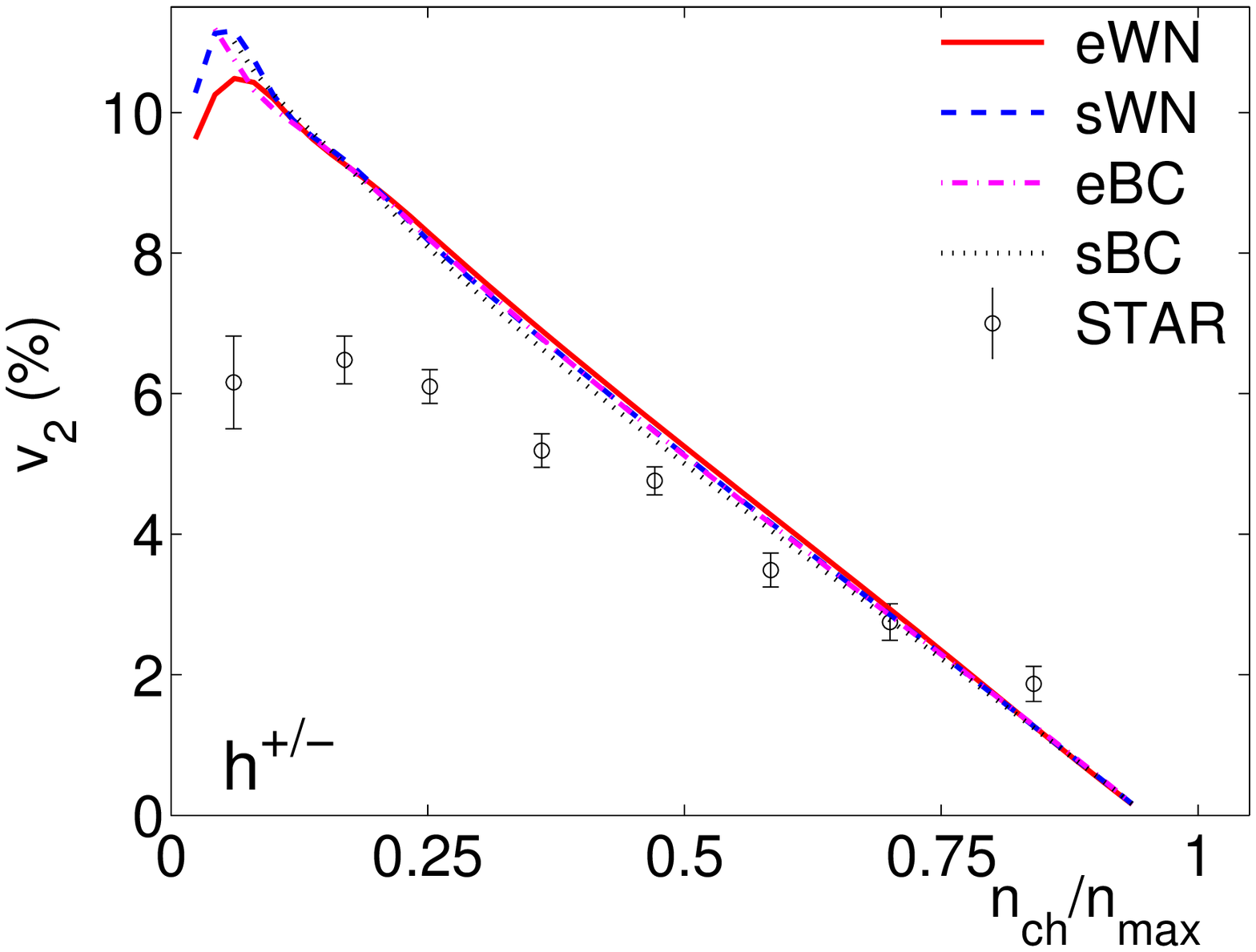}
%\caption{Good sharp prints should be used and not (distorted) photocopies.}
\end{minipage}
\begin{minipage}[t]{52mm}
\vspace*{-50mm}\hspace*{-2mm}
\includegraphics*[height=50.5mm,width=51mm]{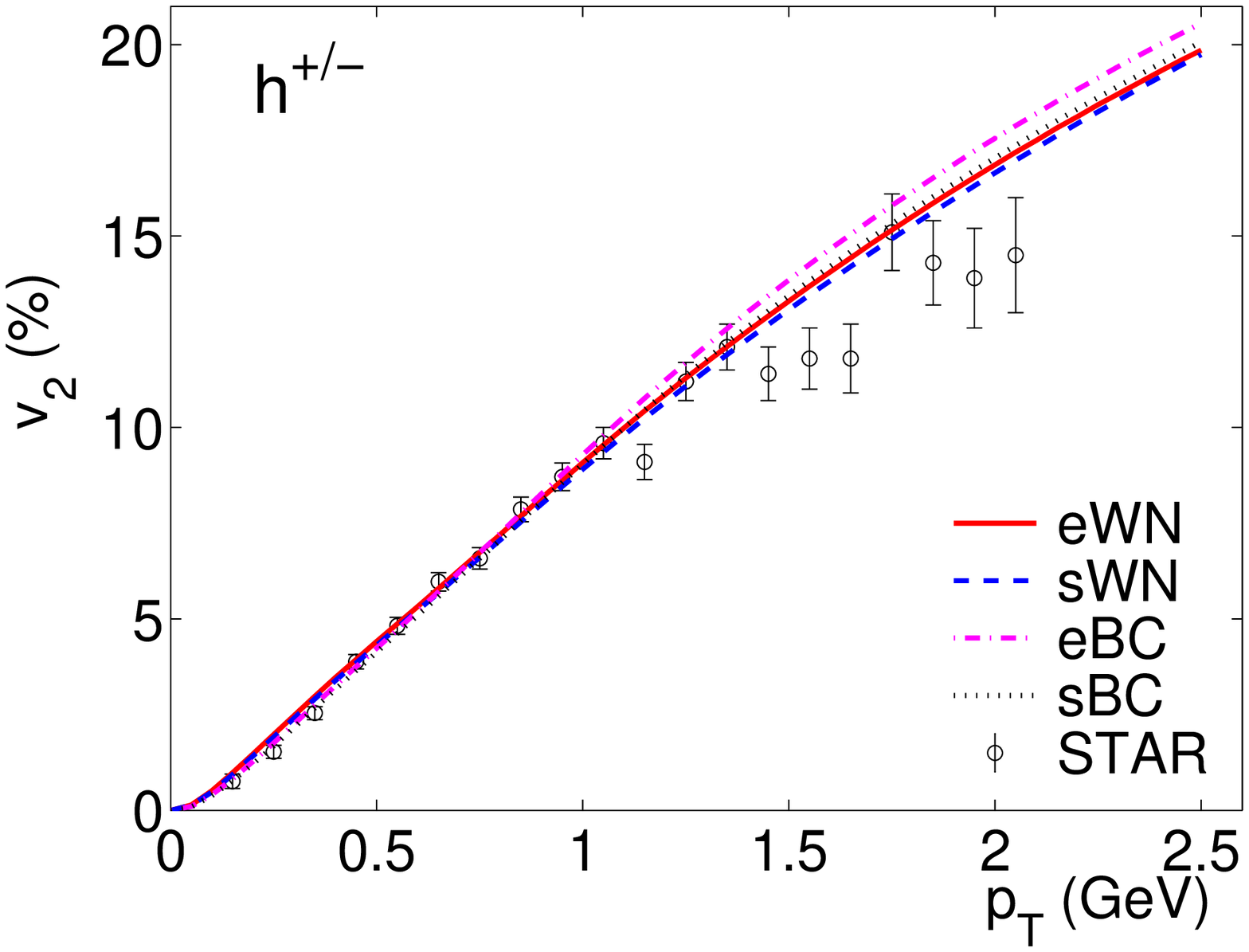}
\end{minipage}
%\hspace{\fill}
%
\begin{minipage}[t]{52mm}
\vspace*{-55mm}\hspace*{-2mm}
\includegraphics*[height=56mm,width=60mm]{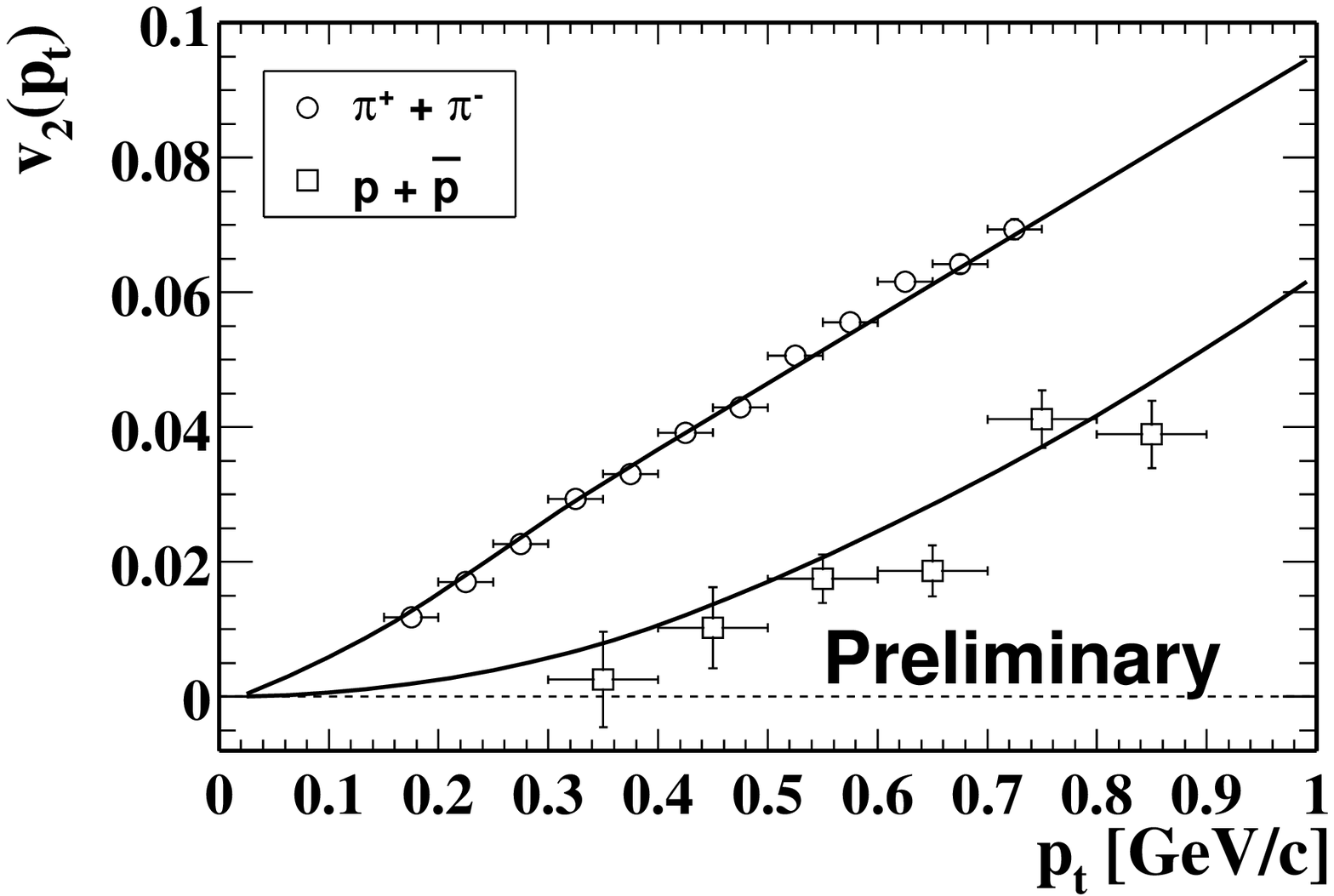}
\end{minipage}
\vspace*{-12mm}
\begin{center}
\begin{minipage}[t]{150mm}
\caption{\label{F2} 
\small 
The elliptic flow coefficient $v_2$ for all charged
particles (left two panels \cite{KHHET}) and for identified pions and 
protons (right panel \cite{Snellings:2001nf}) from 130\,$A$\,GeV Au+Au 
collisions. The left panel shows the $p_\perp$-averaged elliptic flow
as a function of collision centrality, parametrized by the charged 
multiplicity density $n_{\rm ch}$ at midrapidity ($n_{\rm max}$ corresponds 
to the value in central collisions). The right two panels show the 
differential elliptic flow $v_2(p_\perp)$ for minimum bias collisions. 
The data were collected by the STAR collaboration 
\cite{Ackermann:2001tr,Snellings:2001nf,Adler:2001nb}. The curves in the 
left two panels are hydrodynamic calculations corresponding to different 
choices for the initial energy density profile (see \cite{KHHET} for 
details). The curves in the right panel were published in 
\cite{Huovinen:2001cy}.}
\end{minipage}
\end{center}
\end{figure}
%%%%%%%%%%%%%%%%%%%%%%%%%%%%%%%%%%%%%%%%%%%%%%%%%%%%%%%%%%%%%%%%%%%%%%%%%%%%%%%

\vspace*{-10mm}

Figure~\ref{F2} shows the elliptic flow coefficient $v_2$ from Au+Au 
collisions at RHIC \cite{Ackermann:2001tr,Snellings:2001nf,Adler:2001nb}
compared with hydrodynamic calculations. For impact parameters 
$b{\,\leq\,}7$\,fm (corresponding to $n_{\rm ch}/n_{\rm max}{\,\geq\,}0.5$)
and transverse momenta $p_\perp{\,\leq\,}1.5{\,-\,}2$\,GeV/$c$ the data
are seen to exhaust the upper limit for $v_2$ obtained from the hydrodynamic
calculations. For larger impact parameters $b{\,>\,}7$\,fm the 
$p_\perp$-averaged elliptic flow $v_2$ increasingly lags behind the 
hydrodynamic prediction, indicating a lack of early thermalization when 
the initial overlap region becomes too small. The $p_\perp$-differential
elliptic flow stops following the hydrodynamic curves for 
$p_\perp{\,>\,}2$\,GeV/$c$ \cite{Snellings:2001nf} (not shown in 
Fig.~\ref{F2}), indicating incomplete thermalization of high-$p_\perp$ 
particles. Both these effects are expected; what is surprising is the 
excellent agreement otherwise, including the hydrodynamically predicted 
mass-dependence of $v_2$ \cite{Huovinen:2001cy} as seen in the right 
panel of Fig.~\ref{F2}. 

The high level of agreement with hydrodynamics becomes even more 
impressive after you begin to realize how easily it is destroyed: 
As stressed in Sec.~\ref{sec1}, it requires the build-up of momentum
anisotropies during the very early collision stages when the spatial
anisotropy of the reaction zone is still appreciable, causing significant
anisotropies of the pressure gradients. A delay in thermalization by more 
than about 1\,fm/$c$ (2\,fm/$c$) dilutes the spatial anisotropy and the 
hydrodynamically predicted elliptic flow coefficient by 10\% (25\%) 
\cite{Kolb:2000sd} which is more than is allowed by the data. Parton cascade
simulations with standard HIJING input generate almost no elliptic flow
and require an artificial increase of the opacity of the partonic matter
by a factor 80 to reproduce the RHIC data \cite{Molnar:2001ux}. Hadronic
cascades of the RQMD and URQMD type (in which the high-density initial 
state is parametrized by non-interacting, pressureless QCD strings) 
predict \cite{Bleicher:2000sx} too little elliptic flow and a decrease of 
$v_2$ from SPS to RHIC, contrary to the data.

The elliptic flow is self-quenching \cite{Sorge:1997pc}: it makes the 
reaction zone grow faster along its initially short direction and thus
eventually eliminates its own cause. As the spatial deformation of the 
fireball goes to zero, the elliptic flow saturates \cite{Kolb:2000sd}.
The saturation time scale times $c$ is of the order of the transverse 
size of the initial overlap region (at lower energies it is a bit longer, 
see Figs.~7,\,9 in \cite{Kolb:2000sd}). At RHIC energies and above, the
time it takes the collision zone to dilute from the high initial energy 
density to the critical value for hadronization is equal to or longer 
than this saturation time: most or all of the elliptic flow is generated 
before any hadrons even appear! It thus seems that the only possible 
conclusion from the successful hydrodynamic description of the observed 
radial and elliptic flow patterns is that the thermal pressure driving 
the elliptic flow is partonic pressure, and that the early stage of the 
collision must have been a thermalized quark-gluon plasma.     

%%%%%%%%%%%%%%%%%%%%%%%%%%%%%%%%%%%%%%%%%%%%%%%%%%%%%%%%%%%%%%%%%%%%%%%%%%%%%
\section{THE RHIC HBT PUZZLE}
\label{sec4}
%%%%%%%%%%%%%%%%%%%%%%%%%%%%%%%%%%%%%%%%%%%%%%%%%%%%%%%%%%%%%%%%%%%%%%%%%%%%%

Hydrodynamics not only predicts the momentum-space structure of the
hadron emitting source at freeze-out, but also its spatial structure. 
Bose-Einstein (a.k.a. Hanbury Brown-Twiss (HBT)) two-particle intensity 
interferometry allows to access the r.m.s. widths of the space-time 
distribution of hadrons with a given momentum $p$ \cite{Heinz:1999rw}.
One of the interesting questions one can try to address with this tool
is whether at RHIC the reaction zone really flips the sign of its 
spatial deformation between initial impact and final freeze-out, as 
predicted by hydrodynamics \cite{Kolb:2000sd} where the reaction zone
changes from a significant initial elongation {\em perpendicular to} the 
reaction plane to a smaller final elongation {\em into} the reaction 
plane. The answer to this question turns out to be non-trivial, on two
different levels: first, hydrodynamics, at least with the presently 
implemented initial conditions and freeze-out algorithm, fails to
reproduce even for central Au+Au collisions the measured HBT radii 
extracted from two-pion correlations \cite{Adler:2001zd,Johnson:2001zi}. 
I'll show how and explain why. Second, for expanding systems the HBT 
radii don't measure the entire freeze-out region, but only the effective 
emission regions (``regions of homogeneity'') for particles of given 
momentum \cite{Heinz:1999rw}. For non-central collisions, due to the 
anisotropic transverse flow these turn out to have a different spatial 
deformation than the entire (momentum-integrated) freeze-out region, 
giving rise to a different behaviour of the HBT radii observed at 
different angles relative to the reaction plane than perhaps naively 
expected.

%%%%%%%%%%%%%%%%%%%%%%%%% Fig. 3 %%%%%%%%%%%%%%%%%%%%%%%%%%%%%%%%%%%%%%%%%%%%
\vspace*{-5mm}
\begin{figure}[htb]
\begin{center}
%\begin{minipage}[t]{52mm}
\includegraphics*[height=90mm,width=90mm]{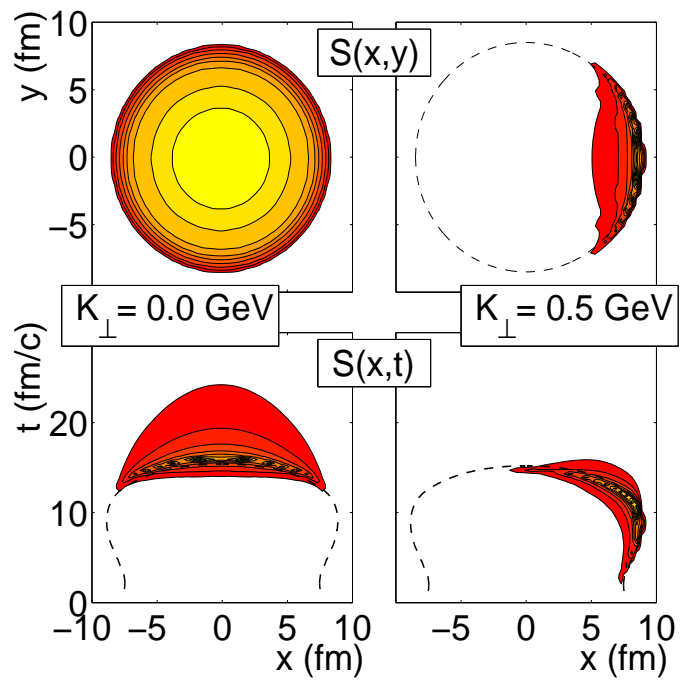}
%\end{minipage}
\end{center}
\begin{center}
\vspace*{-15mm}
\begin{minipage}[t]{150mm}
\caption{\label{F3} 
\small 
Density contours for the effective emission regions of $Y\eq0$ pion 
pairs with transverse momentum $K_\perp\eq0$ (left panels) and 
$K_\perp\eq0.5$\,GeV/$c$ in $x$-direction (right panels). The upper 
and lower panels show projections on the transverse $x$-$y$ plane and on 
the $x$-$t$ plane, respectively.
}
\end{minipage}
\end{center}
\end{figure}
%%%%%%%%%%%%%%%%%%%%%%%%%%%%%%%%%%%%%%%%%%%%%%%%%%%%%%%%%%%%%%%%%%%%%%%%%%%%%%%

\vspace*{-12mm}

\begin{figure}[htb]
\begin{center}
\begin{minipage}[t]{65mm}
\includegraphics*[height=60mm,width=65mm]{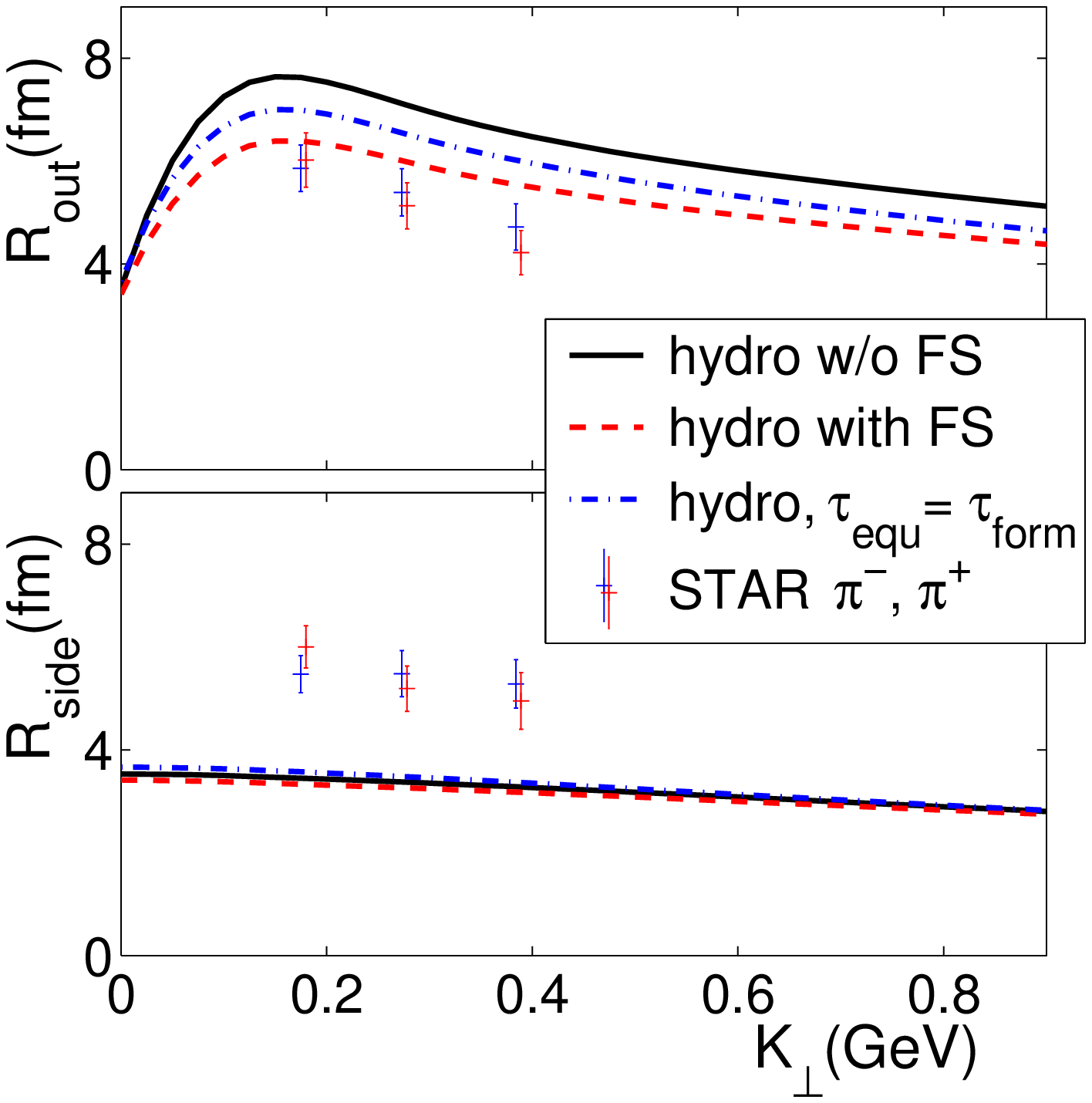}
\end{minipage}
%\hspace{\fill}
%
\begin{minipage}[t]{65mm}
\vspace*{-60mm}
%\hspace*{-3mm}
\includegraphics*[height=61.5mm,width=65mm]{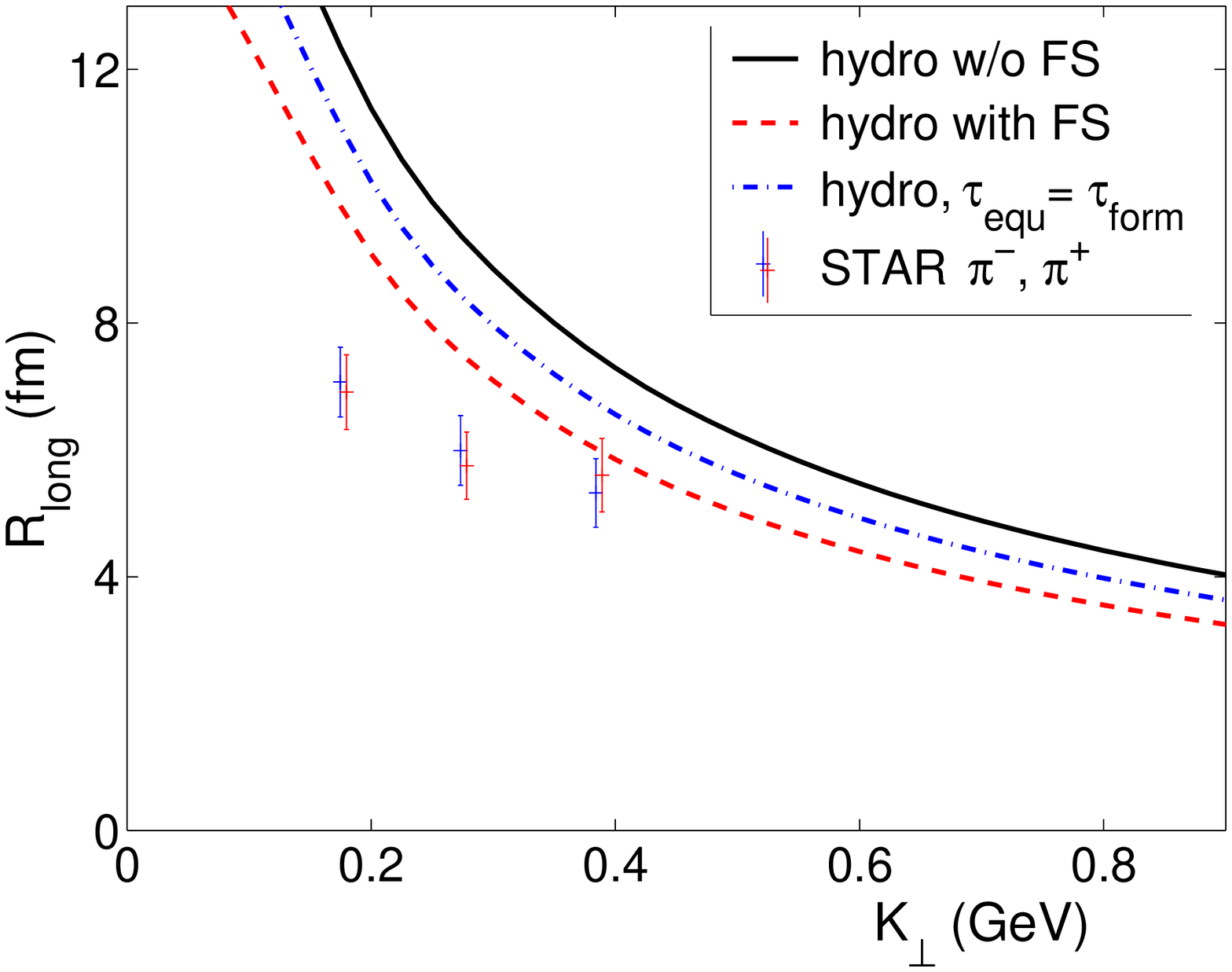}
\end{minipage}
\end{center}
\vspace*{-24mm}
\begin{center}
\begin{minipage}[t]{150mm}
\caption{\label{F4} 
\small 
HBT radii $R_{\rm out},\,R_{\rm side}$, $R_{\rm long}$ for central 
Au+Au collisions at $\sqrt{s}\eq130\,A$\,GeV 
\cite{Adler:2001zd}, compared with hydrodynamic predictions
from the same simulations which provide an excellent fit to the spectra
and elliptic flow (solid lines). See text for the other lines.
}
\end{minipage}
\end{center}
\end{figure}

Figure~\ref{F3} shows the effective emission regions for pion pairs with 
vanishing and nonvanishing transverse momentum $K_\perp$. One sees that
the emission region for pions with $K_\perp\eq0$ is spherically 
symmetric around the fireball center, whereas pions with non-zero 
$K_\perp$ are emitted from a relatively thin crescent-shaped region 
near the edge of the fireball. This apparent ``opacity'' of the source
\cite{Tomasik:1998qt} is a result of the sharp Cooper-Frye freeze-out 
combined with the strong radial flow. It correctly reproduces the 
steeper decrease for increasing $K_\perp\eq0$ of the outward radius 
$R_{\rm out}$ compared with $R_{\rm side}$, which is seen in the
RHIC data \cite{Adler:2001zd,Johnson:2001zi} ({\em cf.} Fig.~\ref{F4}) 
and which was already observed (albeit more weakly) at the SPS 
\cite{Appelshauser:1998rr,Tomasik:1999cq}. 

Unfortunately, Fig.~\ref{F4} also shows that the absolute values of 
$R_{\rm side}$, $R_{\rm out}$, and $R_{\rm long}$ come out quite wrong: 
$R_{\rm side}$ is too small whereas $R_{\rm out}$ and $R_{\rm long}$ 
are both too large when compared with the data. The problem with 
$R_{\rm side}$ from hydrodynamics being too small is well-known from the 
SPS \cite{Schlei:1996mc}; presumably it is mostly due to the sharp 
Cooper-Frye freeze-out and seems to be at least partially resolved if 
the freeze-out kinetics is handled microscopically within a hadronic 
cascade \cite{Soff:2001eh}. The latter gives a more ``fuzzy'' spatial 
freeze-out distribution with larger r.m.s. width in the sideward direction.  
 
%%%%%%%%%%%%%%%%%%%%%%%%% Fig. 4 %%%%%%%%%%%%%%%%%%%%%%%%%%%%%%%%%%%%%%%%%%%%
\vspace*{-8mm}
%%%%%%%%%%%%%%%%%%%%%%%%%%%%%%%%%%%%%%%%%%%%%%%%%%%%%%%%%%%%%%%%%%%%%%%%%%%%%%%

\vspace*{-12mm}

On the other hand, this ``fuzziness'' only exacerbates the problems with
$R_{\rm long}$ and $R_{\rm out}$. It is well known \cite{Heinz:1999rw}
that in high energy heavy-ion collisions the longitudinal HBT radius 
is controlled by the dynamics of the expanding source via the 
longitudinal velocity gradient at freeze-out. For a boost invariant 
longitudinal flow profile this gradient decreases with time as $1/\tau$,
leading to rather weak gradients (and correspondingly large values for
$R_{\rm long}$) at the typical hydrodynamic freeze-out time of
${\sim\,}15$\,fm/$c$. The left lower panel of Fig.~\ref{F3} shows also a long 
emission time duration. This is actually a consequence of the 
large $R_{\rm long}$ because it causes substantial time 
variations along the constant-$\tau$ freeze-out hypersurface within the 
longitudinal homogeneity length. The resulting large value
$(\delta t)^2{\,\equiv\,}\langle (t{-}\bar t)^2\rangle$ for the emission
duration adds to $R_{\rm out}^2\eq\langle(x{-}\bar x)^2
\rangle  + \beta_\perp^2(\delta t)^2 
- 2\beta_\perp\langle(x{-}\bar x)(t{-}\bar t)\rangle$ \cite{Heinz:1999rw} 
(where $\beta_\perp{\eq}K_\perp/K^0$ is the transverse velocity of the 
pion pair) and makes $R_{\rm out}$ come out significantly larger than
$R_{\rm side}$, contrary to the data (see Fig.~\ref{F4}). These problems
are, if anything, worse when freeze-out is handled microscopically 
\cite{Soff:2001eh}.

One possibility to make both $R_{\rm long}$ and $R_{\rm out}$ smaller
would be to force the system to decouple earlier. If there were already
some initial transverse collective motion at the thermalization time when 
the hydrodynamic simulaton is started, this might help to build up 
transverse flow more quickly, leading to an earlier decoupling. We 
have tested this idea \cite{KTH} with two extreme 
assumptions about the transverse expansion prior to thermalization: 
in one simulation, shown as the dot-dashed curve in Fig.~\ref{F4}, we 
started the hydrodynamic evolution directly at the parton formation 
time (for which we took the somewhat arbitrary value $\tau_{\rm
 form}\eq0.2$\,fm/$c$). In another limit (dashed lines in Fig.~\ref{F4}), 
we let the partons stream freely from time $\tau_{\rm form}$ to 
$\tau_{\rm eq}$ and matched at $\tau_{\rm eq}$ the first row of the 
energy momentum tensor to an ideal fluid form, thereby extracting an 
initial transverse flow velocity at $\tau_{\rm eq}$. In both cases the 
resulting transverse flow ``seed'' at $\tau_{\rm eq}\eq0.6$\,fm/$c$ 
caused the system to expand more rapidly and farther out into the 
transverse direction, freezing out 10-20\% earlier.
Fig.~\ref{F4} shows that this helps with both $R_{\rm long}$ and 
$R_{\rm out}$, but not as much as required by the data. And even though
the system expanded to larger values of $r$, $R_{\rm side}$ wouldn't
grow (see Fig.~\ref{F4}) because the homogeneity region only moved 
farther out but its size did not increase. We don't see a way to move
closer to the data by further modifying the initial conditions, and even
the alterations we made to obtain Fig.~\ref{F4} may turn out to be
excluded by the singles spectra and the elliptic flow data (which we 
haven't tested yet). We therefore believe that a resolution to the HBT
puzzle must lie in the handling of the freeze-out process (although we do
not know yet how). 

%%%%%%%%%%%%%%%%%%%%%%%%% Fig. 5 %%%%%%%%%%%%%%%%%%%%%%%%%%%%%%%%%%%%%%%%%%%%
\vspace*{-5mm}
\begin{figure}[htb]
\begin{center}
\begin{minipage}[t]{72mm}
\includegraphics*[height=50mm,width=70mm]{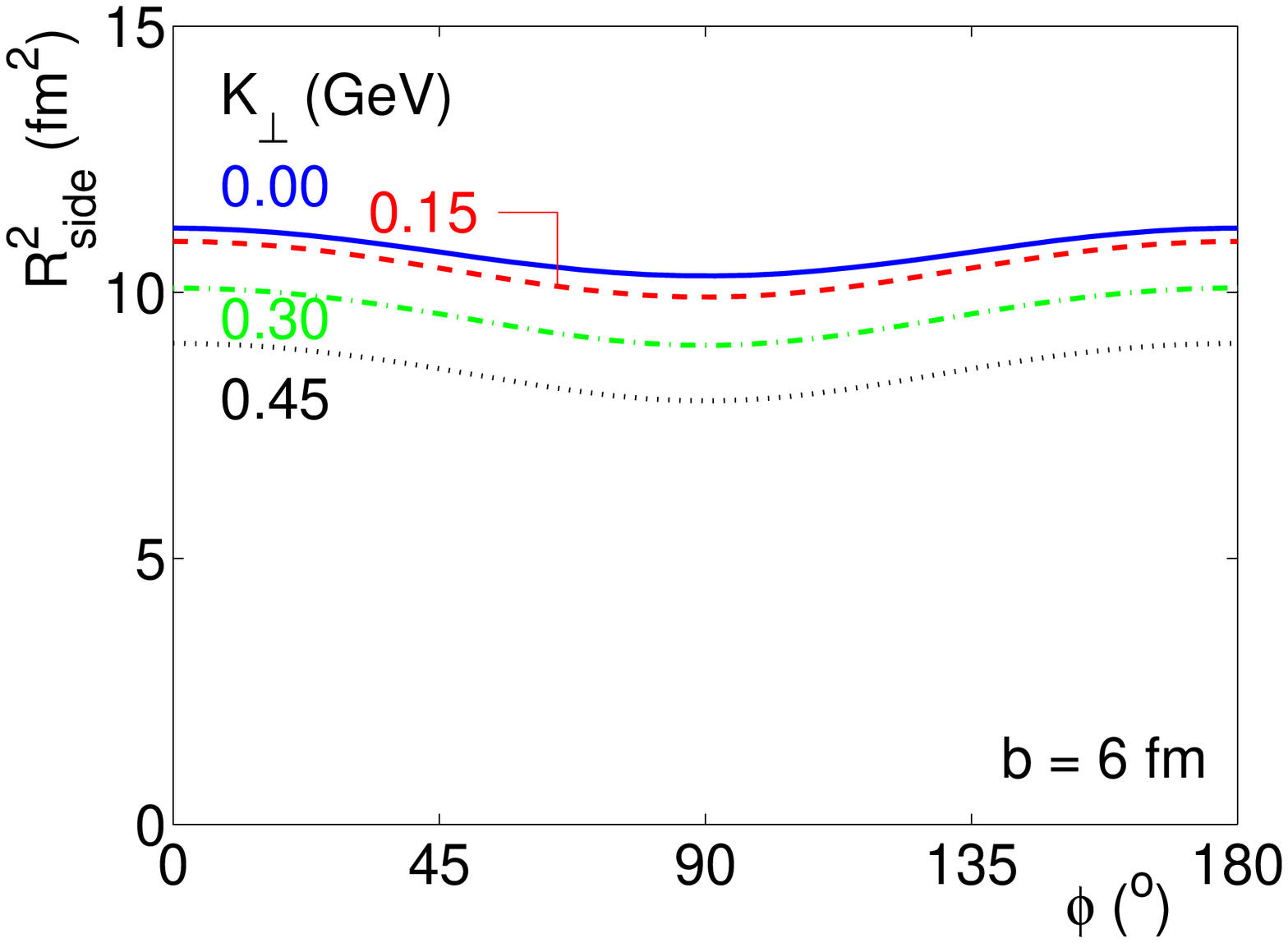}
\end{minipage}
\begin{minipage}[t]{72mm}
\includegraphics*[height=50mm,width=70mm]{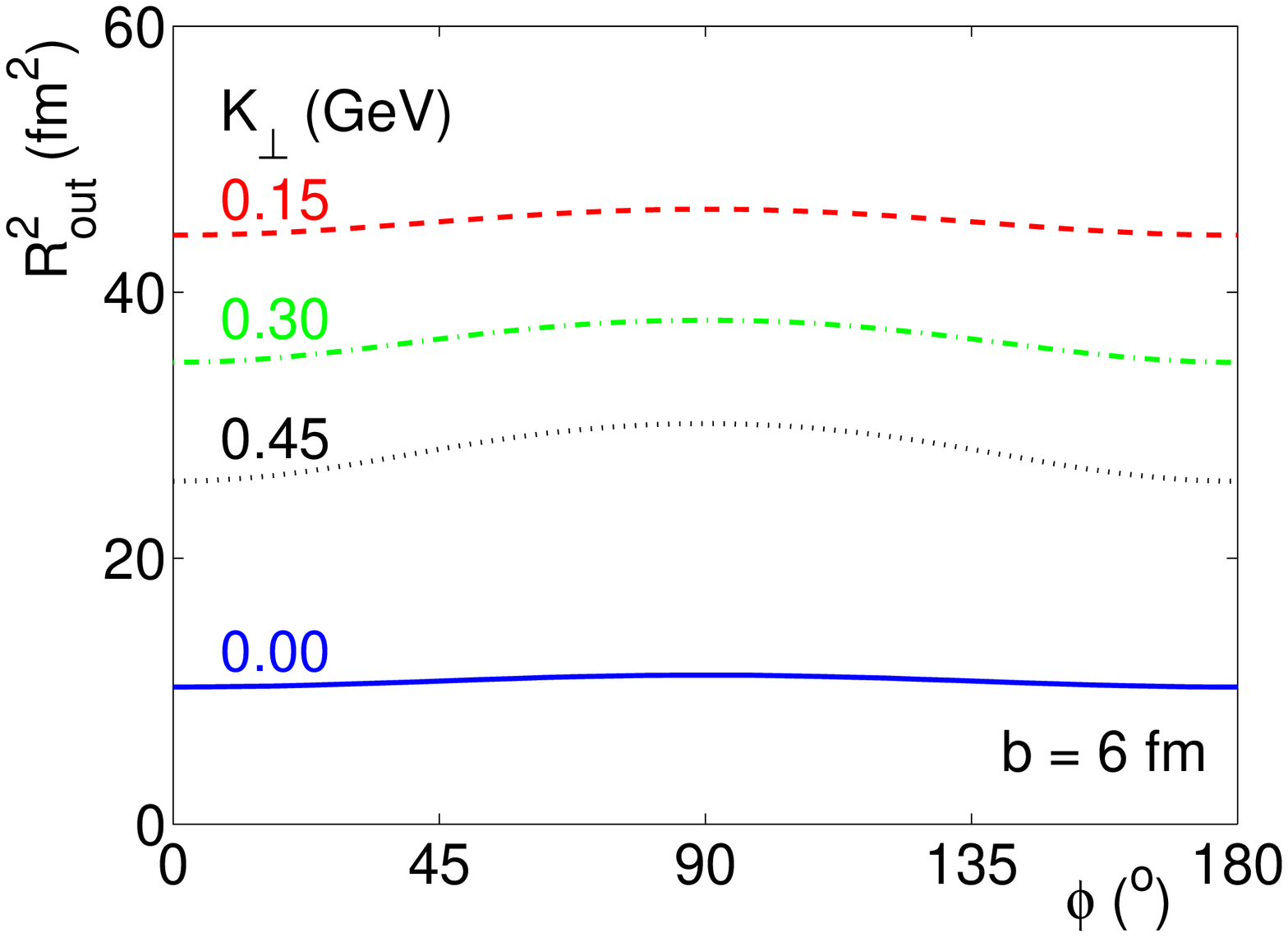}
\end{minipage}
\\
\begin{minipage}[t]{72mm}
\hspace*{-0.5mm}
\includegraphics*[height=50mm,width=70.7mm]{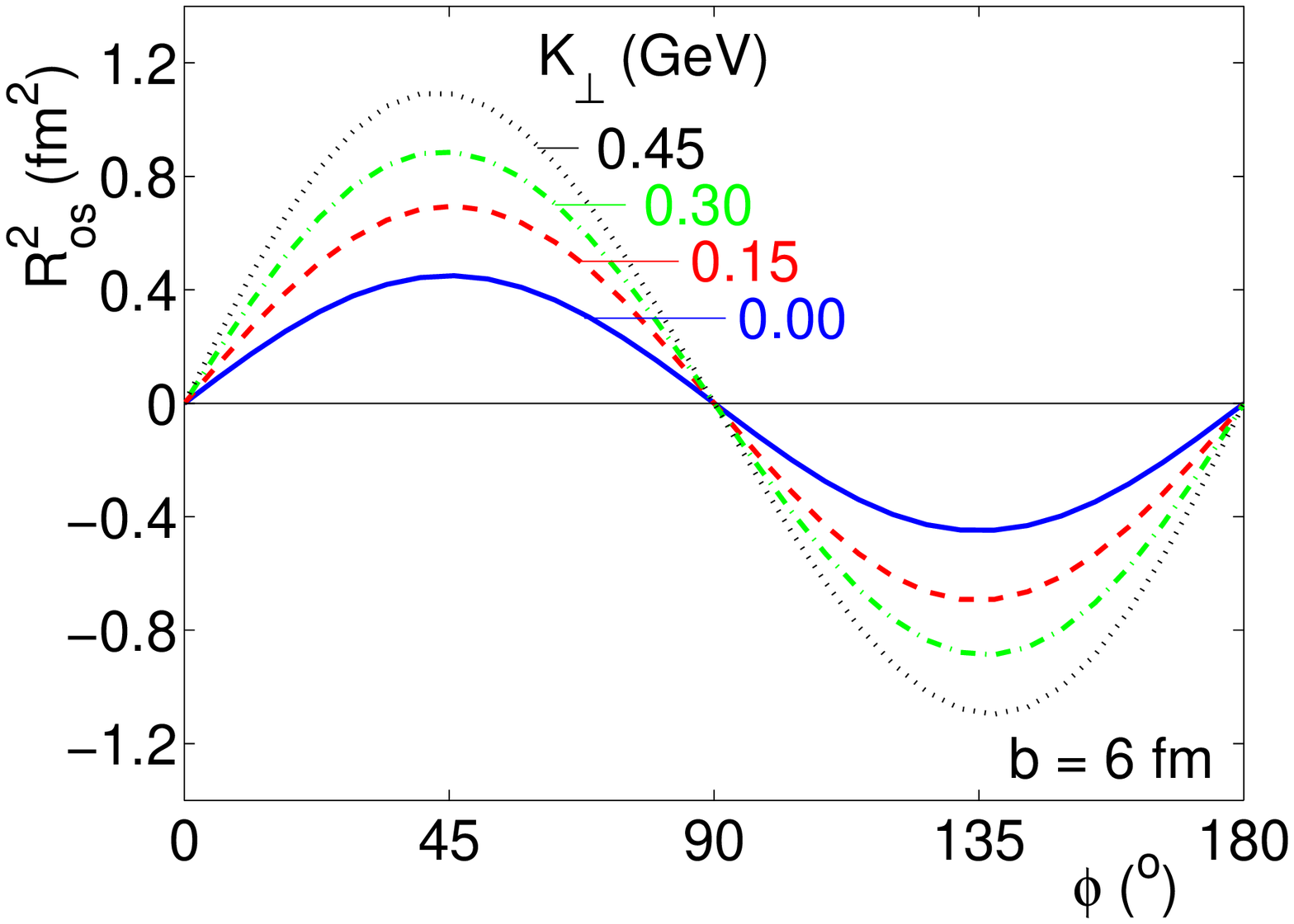}
\end{minipage}
\begin{minipage}[t]{75mm}
\hspace*{-2mm}
\includegraphics*[height=50mm,width=71.9mm]{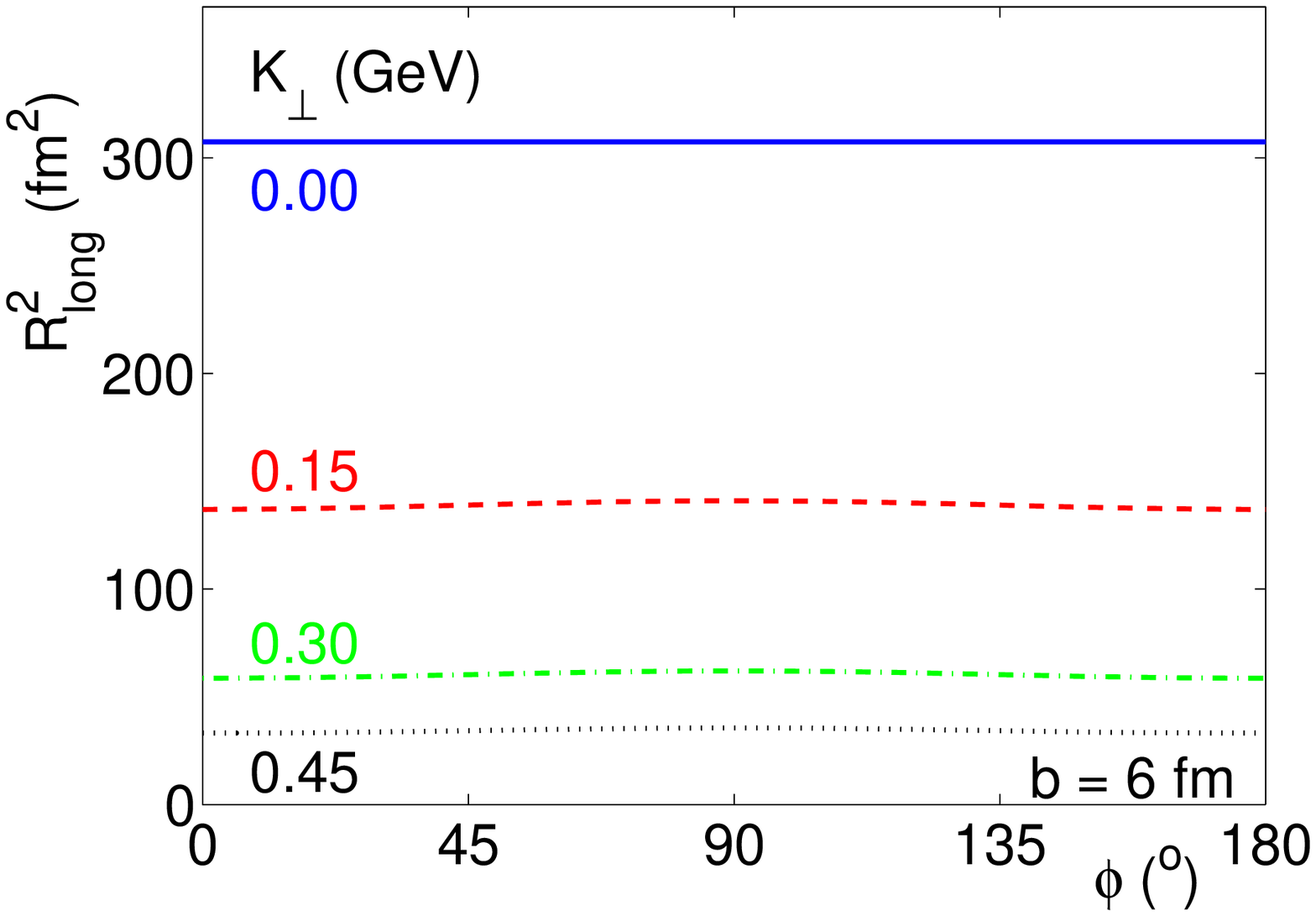}
\end{minipage}
\end{center}
\begin{center}
\vspace*{-15mm}
\begin{minipage}[t]{150mm}
\caption{\label{F5} 
\small 
HBT radius parameters for 130\,$A$\,GeV Au+Au collisions at $b\eq6$\,fm
as a function of the angle $\Phi$ of $\vec K_\perp$ with respect to
the reaction plane, from hydrodynamics.
}
\end{minipage}
\end{center}
\end{figure}
%%%%%%%%%%%%%%%%%%%%%%%%%%%%%%%%%%%%%%%%%%%%%%%%%%%%%%%%%%%%%%%%%%%%%%%%%%%%%%%

\vspace*{-12mm}

Let me close with a quick preview of results for the HBT radii for 
non-central collisions, in particular their dependence on the angle 
$\Phi$ of the transverse pair momentum $\vec K_\perp$ relative to the
reaction plane \cite{Wiedemann:1998cr}. Fig.~\ref{F5} shows hydrodynamic 
results for $R_{\rm side}^2$, $R_{\rm out}^2$, $R_{\rm os}^2$, and 
$R_{\rm long}^2$, plotted as functions of $\Phi$ for a number of 
different values of $K_\perp$. (Note that
``out'' and ``side'' denote the directions parallel and perpendicular 
to $\vec K_\perp$ in the transverse plane \cite{Heinz:1999rw}.) While
$R_{\rm long}^2$ is almost independent of $\Phi$, the three other radius
parameters show marked azimuthal dependences of the generic form (with 
all coefficients being positive)
 \begin{eqnarray}
   R_{\rm side}^2(\Phi) = R_{{\rm s},0}^2 + R_{{\rm s},2}^2 \cos(2\Phi),\ \
   R_{\rm out}^2(\Phi) = R_{{\rm o},0}^2 - R_{{\rm o},2}^2 \cos(2\Phi),\ \
   R_{\rm os}^2(\Phi) = R_{{\rm os},2}^2 \sin(2\Phi).
 \nonumber
 \end{eqnarray}
Although the magnitudes of the coefficients $R_\alpha^2$ in Fig.~\ref{F5} 
are quite different (and presumably not too trustworthy, given the 
disagreement with the data for central collisions in Fig.~\ref{F4}), 
it is surprising that the signs and phases
of the oscillations are identical to those calculated and measured
at the AGS \cite{Lisa:2000ip}! At the AGS radial flow effects are thought 
to be sufficiently weak that the oscillations can be interpreted purely
geometrically \cite{Lisa:2000ip}, reflecting a spatially deformed source 
which is elongated perpendicular to the reaction plane (as is the case for 
the initial overlap region). But hydrodynamics predicts a freeze-out
configuration which is slightly longer {\em in} the reaction plane, as a
result of a much stronger expansion in this direction than in the 
perpendicular one. Shouldn't the oscillations then have different phases?

The answer is: not necessarily. Due to the strong radial flow at RHIC, 
the effective emission regions for pairs of given momentum $K_\perp$ 
(as measured by the HBT correlations) cover only a fraction of the 
transverse source area. Fig.~\ref{F3} shows that for sufficiently large 
$K_\perp$, these effective emission regions are pushed out of the 
fireball center and ``squashed'' towards the transverse edges of the 
source where one has the largest transverse flow velocities. This effect 
increases with the transverse flow and with $K_\perp$. Flow gradients 
reduce the regions of homogeneity \cite{Heinz:1999rw}, leading to the 
shortest correlation radii in the direction with the fastest expansion. 

%%%%%%%%%%%%%%%%%%%%%%%%% Fig. 6 %%%%%%%%%%%%%%%%%%%%%%%%%%%%%%%%%%%%%%%%%%%%
\vspace*{-8mm}
\begin{figure}[htb]
\begin{center}
%\begin{minipage}[t]{52mm}
\includegraphics[height=80mm,width=110mm]{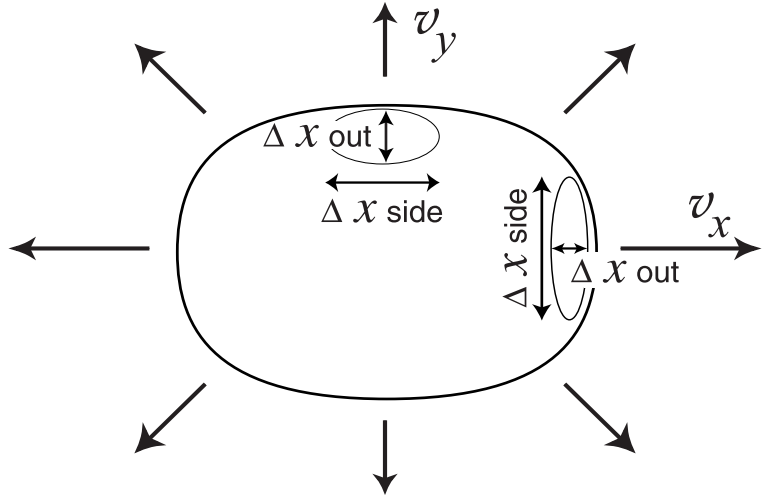}
%\end{minipage}
\end{center}
\begin{center}
\vspace*{-22mm}
\begin{minipage}[t]{150mm}
\caption{\label{F6} 
\small 
Schematic plot of the effective emission regions in the fireball created
in pe\-ri\-pheral Au+Au collisions at RHIC, as predicted by hydrodynamics.
}
\end{minipage}
\end{center}
\end{figure}
%%%%%%%%%%%%%%%%%%%%%%%%%%%%%%%%%%%%%%%%%%%%%%%%%%%%%%%%%%%%%%%%%%%%%%%%%%%%%%%

\vspace*{-12mm}

Fig.~\ref{F6} shows that if the transverse flow is stronger in the 
reaction plane than perpendicular to it ($v_x{\,>\,}v_y$), then for 
pairs emitted in $x$-direction the ``outward'' width $\Delta x_{\rm out}$ 
is {\em smaller} than for pairs emitted in the $y$-direction. Conversely, 
the ``sideward'' width $\Delta x_{\rm side}$ is {\em larger} for pairs 
emitted in $x$-direction, since for them the flow gradients in the 
$x_{\rm side}$-direction are weaker than for pairs emitted in the 
$y$-direction. Hence we have (at least for large $K_\perp$) 
$\Delta x_{\rm out}(\Phi{=}0^\circ){\,<\,}\Delta x_{\rm out}(\Phi{=}90^\circ)$
and 
$\Delta x_{\rm side}(\Phi{=}0^\circ){\,>\,}\Delta x_{\rm
 side}(\Phi{=}90^\circ)$. For small $K_\perp{\,\approx\,}0$ the effective 
emission region is centered at $x{=}y{=}0$, but the stronger flow gradients 
in $x$-direction lead in this case to $\Delta x{\,<\,}\Delta y$ which,
after translation into $\Delta x_{\rm side}$ and $\Delta x_{\rm out}$
at $\Phi{=}0$ and $90^\circ$, gives again rise to the same ordering.
Assuming that the contributions to $R_{\rm out}^2$ involving the emission 
time don't depend strongly on $\Phi$, this explains the sign of the 
oscillations in the upper row of Fig.~\ref{F5}. In the cross term 
$R_{\rm os}^2$, the positive sign of the coefficient multiplying 
$\sin(2\Phi)$ indicates that the major axes $(x_1,x_2)$ of the emission 
ellipsoid are tilted {\em clockwise} relative to the
$(x_{\rm out},x_{\rm side})$ axes in the $1^{\rm st}$ and
$3^{\rm rd}$ quadrant and {\em counterclockwise} in the $2^{\rm nd}$
and $4^{\rm th}$ quadrant. This is again consistent with the flow picture,
since the stronger flow in $x$-direction compresses the emission region 
more strongly in the $x$- than in the $y$-direction, leading to exactly 
such a tilt.
 
Hydrodynamics thus makes a clear prediction: at RHIC energies, the 
phases of the oscillations of the HBT radii as functions of the 
azimuthal angle $\Phi$ are completely dominated by the anisotropic flow 
pattern; dynamics rules over geometry. Even if the magnitudes of the 
radii so far do not fully agree with the data, this qualitative prediction 
seems quite robust. It will be interesting to see it confirmed (or 
contradicted) by the experiments. 

%%%%%%%%%%%%%%%%%%%%%%%%%%%%%%%%%%%%%%%%%%%%%%%%%%%%%%%%%%%%%%%%%%%%%%%%%%%%%
\section{EPILOGUE}
\label{sec5}
%%%%%%%%%%%%%%%%%%%%%%%%%%%%%%%%%%%%%%%%%%%%%%%%%%%%%%%%%%%%%%%%%%%%%%%%%%%%%

I am deeply grateful to Helmut Satz and Frithjof Karsch for the invitation
to speak at this conference. When Helmut held his first meeting on 
Statistical QCD in Bielefeld in 1980 I was still too young to attend: I
had just obtained my Ph.D. with a thesis on heavy-ion collisions at the 
Coulomb barrier, and I was on my way to my first postdoctoral period in the 
U.S., eager to learn quantum field theory and to work my way up in energy.
But I came here two years later, to Helmut's second Bielefeld meeting, and 
gave my first talk on quark-gluon transport theory for relativistic heavy-ion
collisions. Although Helmut and I never published a paper together, our
interactions have always been strong: he influenced and stimulated me, and 
he supported and challenged me. Sometimes I had to work hard until my 
arguments were sharp enough to convince him; and sometimes this never 
happened. My choice of physics problems was affected by my interactions 
with him, personally and through the literature, and without him my 
physics career would very likely have evolved quite differently. His 
famous crystal clear (sometimes I felt: too clear!) presentations of 
many issues in heavy-ion physics inspired and challenged me. I am sure 
that I am not the only one who can say this of Helmut, and to express 
this loudly and clearly is perhaps the greatest compliment we can pay him. 
Thanks, Helmut, and all my best wishes for a very active career as 
professor emeritus!   

%%% References %%%%%%%%%%%%%%%%%%%%%%%%%%%%%%%%%%%%%%%%%%%%%%%%%%%%%%%%%%%%%%


\begin{thebibliography}{99}

\bibitem{O92} 
  J.-Y.~Ollitrault, Phys.~Rev.~D 46 (1992) 229.

\bibitem{VZ96}
  S.~A.~Voloshin, Y.~Zhang, Z.~Phys. C~70 (1996) 665.

%\cite{Stocker:1986ci}
\bibitem{Stocker:1986ci}
H.~St\"ocker and W.~Greiner,
%``High-Energy Heavy Ion Collisions: Probing The Equation Of State Of Highly Excited Hadronic Matter,''
Phys. Rept. 137 (1986) 277.
%%CITATION = PRPLC,137,277;%%
 
\bibitem{Sh00}
  E.~V. Shuryak, Phys. Rev. C 61 (2000) 034905.

\bibitem{Li00}
  Bao-An Li, Phys. Rev. C 61 (2000) 021903(R).

%\cite{Kolb:2000sd}
\bibitem{Kolb:2000sd}
P.~F.~Kolb, J.~Sollfrank and U.~Heinz,
%``Anisotropic transverse flow and the quark-hadron phase transition,''
Phys.\ Rev.\ C~62 (2000) 054909.
%[arXiv:hep-ph/0006129].
%%CITATION = HEP-PH 0006129;%%

%\cite{Zhang:1999rs}
\bibitem{Zhang:1999rs}
B.~Zhang, M.~Gyulassy, and C.~M.~Ko,
%``Elliptic flow from a parton cascade,''
Phys.\ Lett.\ B 455 (1999) 45.
% [arXiv:nucl-th/9902016].
%%CITATION = NUCL-TH 9902016;%%

%\cite{Molnar:2001ux}
\bibitem{Molnar:2001ux}
D.~Molnar and M.~Gyulassy,
%``Saturation of elliptic flow at RHIC: Results from the covariant elastic  parton cascade model MPC,''
nucl-th/0104073.
%%CITATION = NUCL-TH 0104073;%%

%\cite{Sorge:1997pc}
\bibitem{Sorge:1997pc}
H.~Sorge,
%``Elliptical flow: A signature for early pressure in ultrarelativistic  nucleus nucleus collisions,''
Phys.\ Rev.\ Lett.\  78 (1997) 2309; 
% [arXiv:nucl-th/9610026].
%%CITATION = NUCL-TH 9610026;%%
%\cite{Sorge:1999mk}
%\bibitem{Sorge:1999mk}
% H.~Sorge,
%``Highly sensitive centrality dependence of elliptic flow: A novel  signature of the phase transition in QCD,''
{\em ibid.}  82 (1999) 2048.
% [arXiv:nucl-th/9812057].
%%CITATION = NUCL-TH 9812057;%%

%\cite{Voloshin:2000gs}
\bibitem{Voloshin:2000gs}
S.~A.~Voloshin and A.~M.~Poskanzer,
%``The physics of the centrality dependence of elliptic flow,''
Phys.\ Lett.\ 474 (2000) 27.
% [arXiv:nucl-th/9906075].
%%CITATION = NUCL-TH 9906075;%%

%\cite{Ackermann:2001tr}
\bibitem{Ackermann:2001tr}
K.~H.~Ackermann {\it et al.}  [STAR Collaboration],
%``Elliptic flow in Au + Au collisions at s(N N)**(1/2) = 130-GeV,''
Phys.\ Rev.\ Lett.\  86 (2001) 402.
% [arXiv:nucl-ex/0009011].
%%CITATION = NUCL-EX 0009011;%%

%\cite{Lacey:2001va}
\bibitem{Lacey:2001va}
R.~A.~Lacey {\it et al.} [PHENIX Collaboration], in ``Quark Matter 2001'',
Nucl. Phys. A, in press
%``Elliptic flow measurements with the PHENIX detector,''
[nucl-ex/0105003].
%%CITATION = NUCL-EX 0105003;%%

%\cite{Poskanzer:2001cx}
\bibitem{Poskanzer:2001cx}
For a recent compilation of elliptic flow data see 
A.~M.~Poskanzer,
%``Anisotropic flow at the SPS and RHIC,''
nucl-ex/0110013.
%%CITATION = NUCL-EX 0110013;%%

%\cite{Kolb:1999it}
\bibitem{Kolb:1999it}
P.~F.~Kolb, J.~Sollfrank, and U.~Heinz,
%``Elliptic and hexadecupole flow from AGS to LHC energies,''
Phys.\ Lett.\ B 459 (1999) 667.
%[arXiv:nucl-th/9906003].
%%CITATION = NUCL-TH 9906003;%%

%\cite{Teaney:2001cw}
\bibitem{Teaney:2001cw}
D.~Teaney, J.~Lauret, and E.~V.~Shuryak,
%``Flow at the SPS and RHIC as a quark gluon plasma signature,''
Phys.\ Rev.\ Lett.\ 86 (2001) 4783;
%[arXiv:nucl-th/0011058].
%%CITATION = NUCL-TH 0011058;%%
%\cite{Teaney:2001av}
%\bibitem{Teaney:2001av}
%D.~Teaney, J.~Lauret, and E.~V.~Shuryak,
%``A hydrodynamic description of heavy ion collisions at the SPS and RHIC,''
and nucl-th/0110037.
%%CITATION = NUCL-TH 0110037;%%

%\cite{Kolb:2001fh}
\bibitem{Kolb:2001fh}
P.~F.~Kolb, P.~Huovinen, U.~Heinz, and H.~Heiselberg,
%``Elliptic flow at SPS and RHIC: From kinetic transport to  hydrodynamics,''
Phys.\ Lett.\ B 500 (2001) 232.
%[arXiv:hep-ph/0012137].
%%CITATION = HEP-PH 0012137;%%

%\cite{Huovinen:2001cy}
\bibitem{Huovinen:2001cy}
P.~Huovinen, P.~F.~Kolb, U.~Heinz, P.~V.~Ruuskanen, and S.~A.~Voloshin,
%``Radial and elliptic flow at RHIC: Further predictions,''
Phys.\ Lett.\ B 503 (2001) 58.
%[arXiv:hep-ph/0101136].
%%CITATION = HEP-PH 0101136;%%

%\cite{Adler:2001zd}
\bibitem{Adler:2001zd}
C.~Adler {\it et al.} [STAR Collaboration],
%``Pion interferometry of s(NN)**(1/2) = 130-GeV Au + Au collisions at  RHIC,''
Phys.\ Rev.\ Lett.\  87 (2001) 082301.
%[arXiv:nucl-ex/0107008].
%%CITATION = NUCL-EX 0107008;%%

%\cite{Johnson:2001zi}
\bibitem{Johnson:2001zi}
S.~C.~Johnson [PHENIX Collaboration], in ``Quark Matter 2001'', 
Nucl. Phys. A, in press 
%``First measurements of pion correlations by the PHENIX experiment,''
[nucl-ex/0104020].
%%CITATION = NUCL-EX 0104020;%%

%\cite{Rischke:1998fq}
\bibitem{Rischke:1998fq}
D.~H.~Rischke, in ``Hadrons in Dense Matter and Hadrosynthesis'',
J. Cleymans {\it et al.} (eds.), Springer Lecture Notes in Physics Vol. 516,
p. 21, Springer Verlag, Heidelberg, 1999 [nucl-th/9809044]. 
%``Fluid dynamics for relativistic nuclear collisions,''
%arXiv:nucl-th/9809044.
%%CITATION = NUCL-TH 9809044;%%

%\cite{Cooper:1974mv}
\bibitem{Cooper:1974mv}
F.~Cooper and G.~Frye,
%``Comment On The Single Particle Distribution In The Hydrodynamic And Statistical Thermodynamic Models Of Multiparticle Production,''
Phys.\ Rev.\ D 10 (1974) 186.
%%CITATION = PHRVA,D10,186;%%

%\cite{Schnedermann:1994gc}
\bibitem{Schnedermann:1994gc}
E.~Schnedermann and U.~Heinz,
%``A Hydrodynamical assessment of 200-A/GeV collisions,''
Phys.\ Rev.\ C 50 (1994) 1675.
%[arXiv:nucl-th/9402018].
%%CITATION = NUCL-TH 9402018;%%

%\cite{Bass:2000ib}
\bibitem{Bass:2000ib}
S.~A.~Bass and A.~Dumitru,
%``Dynamics of hot bulk QCD matter: From the quark-gluon plasma to  hadronic freeze-out,''
Phys.\ Rev.\ C 61 (2000) 064909.
%[arXiv:nucl-th/0001033].
%%CITATION = NUCL-TH 0001033;%%

%\cite{Heinz:1999kb}
\bibitem{Heinz:1999kb}
U.~Heinz,
%``Primordial hadrosynthesis in the little bang,''
Nucl.\ Phys.\ A 661 (1999) 140c.
%[arXiv:nucl-th/9907060].
%%CITATION = NUCL-TH 9907060;%%

%\cite{Bjorken:1983qr}
\bibitem{Bjorken:1983qr}
J.~D.~Bjorken,
%``Highly Relativistic Nucleus-Nucleus Collisions: The Central Rapidity Region,''
Phys.\ Rev.\ D 27 (1983) 140.
%%CITATION = PHRVA,D27,140;%%

\bibitem{KHHET}
  P.~F.~Kolb, U. Heinz, P. Huovinen, K.~J. Eskola, and K. Tuominen, 
  Nucl. Phys. A 696 (2001) 175.

\bibitem{PHOBOS}
  B.~B.~Back {\it et al.} [PHOBOS Collaboration], Phys. Rev. Lett.
  85 (2000) 3100; nucl-ex/0105011; and nucl-ex/0108009.

\bibitem{PHENIX}
  K.~Adcox {\it et al.} [PHENIX Collaboration], Phys. Rev. Lett.
  86 (2001) 3500.

\bibitem{PHENIX_spec}
  J. Velkovska {\it et al.} [PHENIX Collaboration], nucl-ex/0105012.

\bibitem{STAR_spec}
  J.~Harris {\it et al.} [STAR Collaboration], in ``Quark Matter
  2001'', Nucl. Phys. A, in press; C. Adler {\it et al.} [STAR Collaboration],
  nucl-ex/0110009.

%\cite{Lee:1990sk}
\bibitem{Lee:1990sk}
K.~S.~Lee, U.~Heinz, and E.~Schnedermann,
%``Search For Collective Transverse Flow Using Particle Transverse Momentum Spectra In Relativistic Heavy Ion Collisions,''
Z.\ Phys.\ C 48 (1990) 525.
%%CITATION = ZEPYA,C48,525;%%

%\cite{Braun-Munzinger:2001ip}
\bibitem{Braun-Munzinger:2001ip}
P.~Braun-Munzinger, D.~Magestro, K.~Redlich, and J.~Stachel,
%``Hadron production in Au Au collisions at RHIC,''
Phys.\ Lett.\ B 518 (2001) 41.
%[arXiv:hep-ph/0105229].
%%CITATION = HEP-PH 0105229;%%

%\cite{Snellings:2001nf}
\bibitem{Snellings:2001nf}
R.~J.~Snellings  [STAR Collaboration],
%``Elliptic flow in Au + Au collisions at s(N N)**1/2 = 130-GeV,''
in ``Quark Matter 2001'', Nucl. Phys. A, in press [nucl-ex/0104006].
%%CITATION = NUCL-EX 0104006;%%

%\cite{Adler:2001nb}
\bibitem{Adler:2001nb}
C.~Adler {\it et al.}  [STAR Collaboration],
%``Identified particle elliptic flow in Au + Au collisions at  s(NN)**(1/2) = 130-GeV,''
Phys.\ Rev.\ Lett.\ 87 (2001) 182301.
%[arXiv:nucl-ex/0107003].
%%CITATION = NUCL-EX 0107003;%%

%\cite{Bleicher:2000sx}
\bibitem{Bleicher:2000sx}
M.~Bleicher and H.~St\"ocker,
%``Anisotropic flow in ultra-relativistic heavy ion collisions,''
hep-ph/0006147.
%arXiv:hep-ph/0006147.
%%CITATION = HEP-PH 0006147;%%

%\cite{Heinz:1999rw}
\bibitem{Heinz:1999rw}
U.~Heinz and B.~V.~Jacak,
%``Two-particle correlations in relativistic heavy-ion collisions,''
Ann.\ Rev.\ Nucl.\ Part.\ Sci.\ 49 (1999) 529;
%[arXiv:nucl-th/9902020].
%%CITATION = NUCL-TH 9902020;%%
%\cite{Wiedemann:1999qn}
%\bibitem{Wiedemann:1999qn}
U.~A.~Wiedemann and U.~Heinz,
%``Particle interferometry for relativistic heavy-ion collisions,''
Phys.\ Rept.\ 319 (1999) 145.
%[arXiv:nucl-th/9901094].
%%CITATION = NUCL-TH 9901094;%%

%\cite{Tomasik:1998qt}
\bibitem{Tomasik:1998qt}
B.~Tomasik and U.~Heinz,
%``Fine-tuning two particle interferometry. II: Opacity effects,''
nucl-th/9805016; and 
%%CITATION = NUCL-TH 9805016;%%
%\cite{Tomasik:1999kz}
%\bibitem{Tomasik:1999kz}
%B.~Tomasik and U.~W.~Heinz,
%``Potentials and limits of Bose-Einstein interferometry,''
Acta Phys.\ Slov.\  49 (1999) 251 [nucl-th/9901006].
%%CITATION = NUCL-TH 9901006;%%

%\cite{Appelshauser:1998rr}
\bibitem{Appelshauser:1998rr}
H.~Appelsh\"auser {\it et al.}  [NA49 Collaboration],
%``Hadronic expansion dynamics in central Pb + Pb collisions at 158-GeV  per nucleon,''
Eur.\ Phys.\ J.\ C 2 (1998) 661.
%[arXiv:hep-ex/9711024].
%%CITATION = HEP-EX 9711024;%%

%\cite{Tomasik:1999cq}
\bibitem{Tomasik:1999cq}
B.~Tomasik, U.~A.~Wiedemann, and U.~Heinz,
%``Reconstructing the freeze-out state in Pb + Pb collisions at  158-A-GeV/c,''
nucl-th/9907096.
%%CITATION = NUCL-TH 9907096;%%

%\cite{Schlei:1996mc}
\bibitem{Schlei:1996mc}
B.~R.~Schlei and N.~Xu,
%``m(T) dependence of Bose-Einstein correlation radii,''
Phys.\ Rev.\ C 54 (1996) 2155;
%[arXiv:nucl-th/9608048].
%%CITATION = NUCL-TH 9608048;%%
%\cite{Schlei:1999zy}
%\bibitem{Schlei:1999zy}
B.~R.~Schlei, D.~Strottman, J.~P.~Sullivan, and H.~W.~van Hecke,
%``Bose-Einstein correlations and the equation of state of nuclear matter,''
Eur.\ Phys.\ J.\ C 10 (1999) 483.
%[arXiv:nucl-th/9809070].
%%CITATION = NUCL-TH 9809070;%%

%\cite{Soff:2001eh}
\bibitem{Soff:2001eh}
S.~Soff, S.~A.~Bass, and A.~Dumitru,
%``Pion interferometry at RHIC: Probing a thermalized quark gluon plasma?,''
Phys.\ Rev.\ Lett.\ 86 (2001) 3981.
%[arXiv:nucl-th/0012085].
%%CITATION = NUCL-TH 0012085;%%

\bibitem{KTH}
  P.~F.~Kolb, M.~Tilley, and U.~Heinz, in preparation. The dashed and
  dot-dashed curves in Fig.~\ref{F4} are preliminary in that they were
  calculated with a somewhat different initial energy density profile
  than the solid line. 

%\cite{Wiedemann:1998cr}
\bibitem{Wiedemann:1998cr}
U.~A.~Wiedemann,
%``Two-particle interferometry for noncentral heavy-ion collisions,''
Phys.\ Rev.\ C 57 (1998) 266.
%[arXiv:nucl-th/9707046].
%%CITATION = NUCL-TH 9707046;%%

%\cite{Lisa:2000ip}
\bibitem{Lisa:2000ip}
M.~A.~Lisa, U.~Heinz, and U.~A.~Wiedemann,
%``Tilted pion sources from azimuthally sensitive HBT interferometry,''
Phys.\ Lett.\ B 489 (2000) 287;
%[arXiv:nucl-th/0003022].
%%CITATION = NUCL-TH 0003022;%%
%\cite{Lisa:2000xj}
%\bibitem{Lisa:2000xj}
M.~A.~Lisa {\it et al.} [E895 Collaboration],
%``Azimuthal dependence of pion interferometry at the AGS,''
Phys.\ Lett.\ B 496 (2000) 1.
%[arXiv:nucl-ex/0007022].
%%CITATION = NUCL-EX 0007022;%%

\end{thebibliography}
\end{document}